\documentclass[aps,showpacs,twocolumn,superscriptaddress,amsmath,amssymb,floatfix,longbibliography]{revtex4-1}

\usepackage{graphicx}
\usepackage{float}
\usepackage{times}
\usepackage{latexsym}
\usepackage{amsmath}
\usepackage{amssymb}
\usepackage{bm}
\usepackage{euscript}
\usepackage{color}
\usepackage{wasysym}
\usepackage{epstopdf}
\usepackage[colorlinks=true,linkcolor=blue,citecolor=blue]{hyperref}
\usepackage{type1cm}
\usepackage{appendix}
\usepackage{subfigure}
\usepackage{multirow}

\usepackage{color}
\definecolor{darkblue}{rgb}{0.,0.,0.4}
\definecolor{darkred}{rgb}{0.5,0.,0.}
\definecolor{BlueViolet}{RGB}{138,43,226}
\definecolor{SkyBlue}{RGB}{30,144,255}
\definecolor{DarkGreen}{RGB}{0,100,0}

\usepackage[normalem]{ulem}

\begin{document}
\title{Eigenstate properties of the disordered Bose-Hubbard chain}

\author{Jie~Chen}
\email{chenjie666@sjtu.edu.cn}
\affiliation{Key Laboratory of Artificial Structures and Quantum Control (Ministry of Education),
School of Physics and Astronomy, Shenyang National Laboratory for Materials Science,
Shanghai Jiao Tong University, Shanghai 200240, China}

\author{Chun~Chen}
\email{chunchen@sjtu.edu.cn}
\affiliation{Key Laboratory of Artificial Structures and Quantum Control (Ministry of Education),
School of Physics and Astronomy, Shenyang National Laboratory for Materials Science,
Shanghai Jiao Tong University, Shanghai 200240, China}

\author{Xiaoqun~Wang}
\email{xiaoqunwang@zju.edu.cn}
\affiliation{Key Laboratory of Artificial Structures and Quantum Control (Ministry of Education),
School of Physics and Astronomy, Shenyang National Laboratory for Materials Science,
Shanghai Jiao Tong University, Shanghai 200240, China}
\affiliation{School of Physics, Zhejiang University, Hangzhou 310058, Zhejiang, China}
\affiliation{Tsung-Dao Lee Institute, Shanghai Jiao Tong University, Shanghai 200240, China}
\affiliation{Collaborative Innovation Center of Advanced Microstructures, Nanjing University, Nanjing 210093, China}

\date{\today}

\begin{abstract}

Many-body localization (MBL) of a disordered interacting boson system in one dimension is studied numerically at the filling faction one-half. The von Neumann entanglement entropy $S_{\textrm{vN}}$ is commonly used to detect the MBL phase transition but remains challenging to be directly measured. Based on the $U(1)$ symmetry from the particle number conservation, $S_{\textrm{vN}}$ can be decomposed into the particle number entropy $S_N$ and the configuration entropy $S_C$. In light of the tendency that the eigenstate's $S_C$ nears zero in the localized phase, we introduce a quantity describing the deviation of $S_N$ from the ideal thermalization distribution; finite-size scaling analysis illustrates that it shares the same phase transition point with $S_{\textrm{vN}}$ but displays the better critical exponents. This observation hints that the phase transition to MBL might largely be determined by $S_N$ and its fluctuations. Notably, the recent experiments [A. Lukin {\it et al}., \href{https://doi.org/10.1126/science.aau0818}{Science {\bf 364}, 256 (2019)}; J. L\'eonard {\it et al}., \href{https://doi.org/10.1038/s41567-022-01887-3}{Nat. Phys. {\bf 19}, 481 (2023)}] demonstrated that this deviation can potentially be measured through the $S_N$ measurement. Furthermore, our investigations reveal that the thermalized states primarily occupy the low-energy section of the spectrum, as indicated by measures of localization length, gap ratio, and energy density distribution. This low-energy spectrum of the Bose model closely resembles the entire spectrum of the Fermi (or spin $X\!X\!Z$) model, accommodating a transition from the thermalized to the localized states. While, owing to the bosonic statistics, the high-energy spectrum of the model allows the formation of distinct clusters of bosons in the random potential background. We analyze the resulting eigenstate properties and briefly summarize the associated dynamics. To distinguish between the phase regions at the low and high energies, a probing quantity based on the structure of $S_{\textrm{vN}}$ is also devised. Our work highlights the importance of symmetry combined with entanglement in the study of MBL. In this regard, for the disordered Heisenberg $X\!X\!Z$ chain, the recent pure eigenvalue analyses in [J. {\v{S}}untajs {\it et al}., \href{https://doi.org/10.1103/PhysRevE.102.062144}{Phys. Rev. E {\bf 102}, 062144 (2020)}] would appear inadequate, while methods used in [A. Morningstar {\it et al}., \href{https://doi.org/10.1103/PhysRevB.105.174205}{Phys. Rev. B {\bf 105}, 174205 (2022)}] that spoil the $U(1)$ symmetry could be misleading.

\end{abstract}

\maketitle

\section{\label{sec:level1}Introduction}

In the absence of many-body interactions, it is known that Anderson localization may occur in a disordered system depending upon dimensionality and disorder strength \cite{1,44,45}. This localization phenomenon, essentially caused by interference of elastic scatterings, has been intensively investigated in various disordered circumstances for more than five decades \cite{46,47,48}. In recent years, with the advances of numerical techniques, computational capacities, and experimental facilities, the interaction effect within many-body disordered systems has become of great interest in light of the novel competition between many-body interactions and quenched disorder. One would like to thoroughly explore the effects of disorder on the fundamental properties of many-body interacting systems associated with ergodicity breaking, eigenstate thermalization, and quantum criticality. Particularly, many-body localization (MBL) is expected to be intrinsically different from Anderson localization \cite{2,53,54,55}.

In the absence of disorder, a nonintegrable interacting system is usually considered to satisfy the so-called eigenstate thermalization hypothesis (ETH) \cite{56,57}, which assumes that the ergodicity of many-body states exists, implying that eigenstate fluctuations in the expectation of a quantum mechanical observable is about the same order of magnitude as the statistical fluctuations in a microcanonical ensemble controlled by an appropriate energy density \cite{2}. When sufficiently strong disorder is introduced into such many-body interacting systems, an ergodic many-body state may turn into an MBL state, in which local integrals of motion (LIOMs) emerge \cite{4,5,6} such that the integrability is considerably restored, resulting in the Poisson distribution of the energy levels, rather than the Wigner-Dyson surmise for the typical ETH phases as is derived from the random matrix theory \cite{2,3}.

Phenomenology of MBL can be relevant for many different but related settings, such as a Floquet disordered system \cite{14,15,16,17}, for which the driven system is considered to implement the so-called discrete time crystal \cite{18,19}, the quantum frustrated Heisenberg spin chains \cite{13}, and the one-dimensional (1D) QED lattice version of the Schwinger model with a disordered gauge \cite{22}, as well as in several open quantum setups \cite{58,59,60,61,62,75}. Physical properties of MBL have been widely studied numerically for 1D Heisenberg or similarly interacting hard-core boson systems with randomness \cite{9,10,11,12,13,29,30}. One finds that by exploiting entanglement entropy, a dynamical transition beyond the quantum phase transition occurs between the ETH and MBL states that is driven by disorder strength when varying the energy density \cite{9,10,11,12,13}. While an ETH state has an extended thermal nature, an MBL state is featured by its insulating nature, being close to a product of localized states. Correspondingly, entanglement entropy possesses a volume law in the ETH phase \cite{13} but an area law in the MBL phase. Moreover, a slowly dephasing process is unveiled in MBL phase, engendering a logarithmical entropy increase upon the time evolution of a many-body state \cite{7,8}.

Many experimental efforts are cast into the measurements of disorder effects for interacting many-body systems \cite{18,19,30,35,49,70,73}. Specifically, A. Lukin {\it et al}.~\cite{35} succeeded in realizing a 1D interacting Aubry-Andr\'e model for bosons. By probing particle number entropy and configuration entropy, they detected that the signature of quantum thermalization, the particle's finite localization length, and an area-law scaling of the particle number entropy can help to distinguish the MBL phase from the ergodic ETH phase. Moreover, they also found a slowly growing behavior of the configuration entropy that ultimately results in a volume-law scaling, demonstrating that the MBL state is qualitatively different from a noninteracting Anderson insulating state. Curiously, the key difference between the Bose model and the Fermi (or spin) model is that Bose statistics can result in the clustering of multiple particles, potentially giving rise to the novel phenomena of Bose type MBL. This distinctive character is missing in \cite{35,leonard2023probing}.

In this paper, we extensively explore the nature of MBL phase in 1D Bose-Hubbard model with disorder by exact diagonalization (ED) \cite{28} and several complementary techniques \cite{37,38}. For the present system, numerical calculations are limited to small sizes since the occupancy number of bosons can be any integer, depending on the filling factor under consideration. However, since Hamiltonian (\ref{dbhmodel}) below commutes with total number of particles, the ED calculation can be performed in the subspace with a given number of particles. For a chain with $L$ sites and $N$ bosons, the dimension for the matrix diagonalization is $(L+N-1)!/[N!(L-1)!] $\cite{28}. By implementing the separation of von Neumann entanglement entropy, one can further compute the particle number entropy and configuration entropy to access the so-called symmetry-resolved entropy. In our numerical calculations, we exclusively focus on the filling factor $1/2$. Considering the rapid growth of the Hilbert-space dimension, it is practically feasible that $10000$ samples were taken for the averages with $L=8$, $5000$ with $L=10$, $3000$ with $L=12$, and for $L=14$, $200$ samples are used for weak disorder, $400$ for medium disorder, and $600$ for strong disorder. We impose open boundary conditions in the calculations to compute the two-body density-density correlation function and periodic boundary conditions for other quantities. To access larger system sizes, we also employ time-evolving block decimation (TEBD) \cite{37,38} method to evaluate the time evolution of the two-body correlation function. Errors resulting from sample averages are given explicitly for most cases and are invisible otherwise. 

In the following, we first introduce the disordered Bose-Hubbard (dBH) model and briefly discuss current status regarding the ground-state studies on this model in Section~\ref{sec:sect2}. In Section~\ref{sec:sect3}, we focus on the phase transition from ETH to MBL; at the beginning of this section, a detailed outline of its content is provided. In Section~\ref{sec:sect4}, we analyze different MBL phases within dBH model through the properties of its eigenstates; this section also starts with a detailed content overview. A summary and pertinent discussions are finally given in Section~\ref{sec:sect6}. For completeness, several corresponding supplementary results are included in Appendixes~\ref{sec:append_Decomp}-\ref{sec:append_Locali}.

\section{\label{sec:level1}Model}
\label{sec:sect2}

The 1D dBH model can be written as
\begin{eqnarray}
\hat H&=&-J\sum_i{(\hat{a}_{i}^{\dag}\hat{a}_{i+1}+\textrm{H.c.})}\nonumber\\
&&+\frac{U}{2}\sum_i{\hat{n}_i(\hat{n}_i-1)}+\mu\sum_i{\delta_i\hat{n}_i}
\label{dbhmodel}
\end{eqnarray}
where $\hat{a}_{i}^{\dag},\hat{a}_i$ are particle creation and annihilation operators, and $\hat{n}_i=\hat{a}_{i}^{\dag}\hat{a}_i$ is the particle occupation number operator at site $i$. Disorder is introduced through a random number $\delta_i$, which is uniformly distributed within $[-1,1]$ for all sites, thus chemical potential $\mu$ becomes a tuning parameter for the disorder strength. For convenience, we set $J=1$ as the energy-scale unit hereafter.

For Hamiltonian~(\ref{dbhmodel}), we use the ED technique to find all its eigenvalues $E_i$ with an order of $E_1\leqslant E_2\leqslant\cdots\leqslant E_D$ and $D$ is the dimension of the Hamiltonian matrix. The energy density $\varepsilon$ used throughout this work is then defined by
\begin{eqnarray}
\varepsilon =\dfrac{E_i-E_1}{E_D-E_1}.
\end{eqnarray}

At zero temperature, the BH model, in the absence of disorder, undergoes a quantum phase transition from a superfluid phase to the Mott-insulator state with an integer filling factor as the interaction strength $U$ increases \cite{39}. When disorder is introduced, the nature of the ground state is anticipated to alter \cite{40,41,42,43}. For a given diagonal disorder sample $\{\delta_i\}$ with $i\in[1, L]$, one finds that the system changes from the superfluid state first to the Bose-glass state and then to the Mott-insulator state under the rise of $U$ \cite{41,42}. In the case of an off-diagonal disorder configuration assigned to $J$, a phase diagram has also been established in terms of the interaction strength $U$ and the randomness strength in the hopping amplitude $J$ \cite{41,42,43}.

In this work, we fix $U=3J$ and always maintain the filling factor to be one-half per site, i.e., $N=\sum_i\hat{n}_i=\frac{L}{2}$, for which one has a superfluid ground state in the case of a clean chain \cite{50}.

\section{\label{sec:level1}Eigenstate Phase Transition}
\label{sec:sect3}

A plethora of physical quantities have been used to numerically detect the MBL phase transition, but most are experimentally inaccessible. Among these, the von Neumann entropy $S_{\textrm{vN}}$ is a common choice, yet remains challenging to be directly measured. Conceiving the experimentally accessible probes to assess the MBL transition therefore comprises one of the main goals of the present work.

\begin{figure}[t]
	\begin{center}
		\includegraphics[scale=0.6]{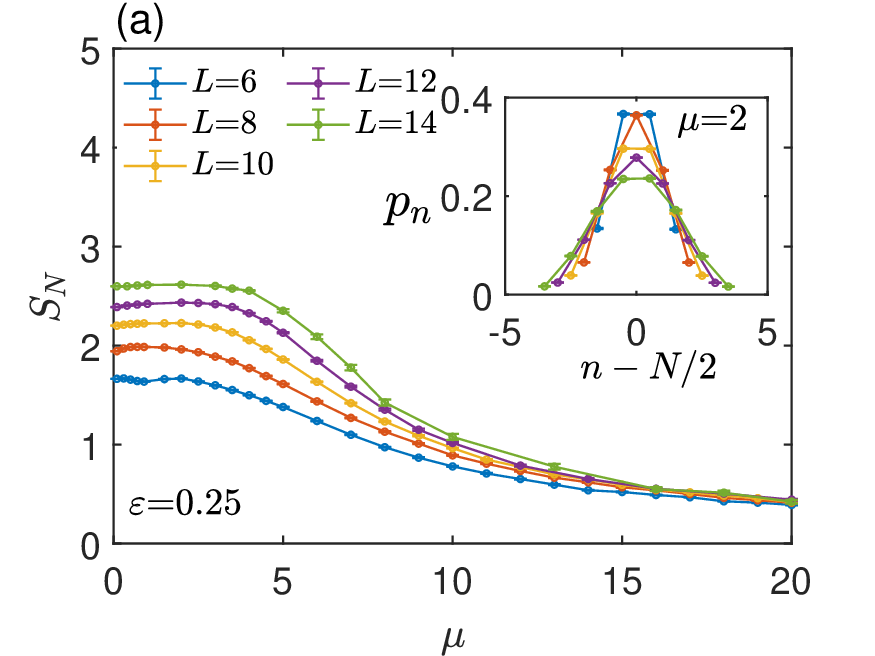}
		\includegraphics[scale=0.6]{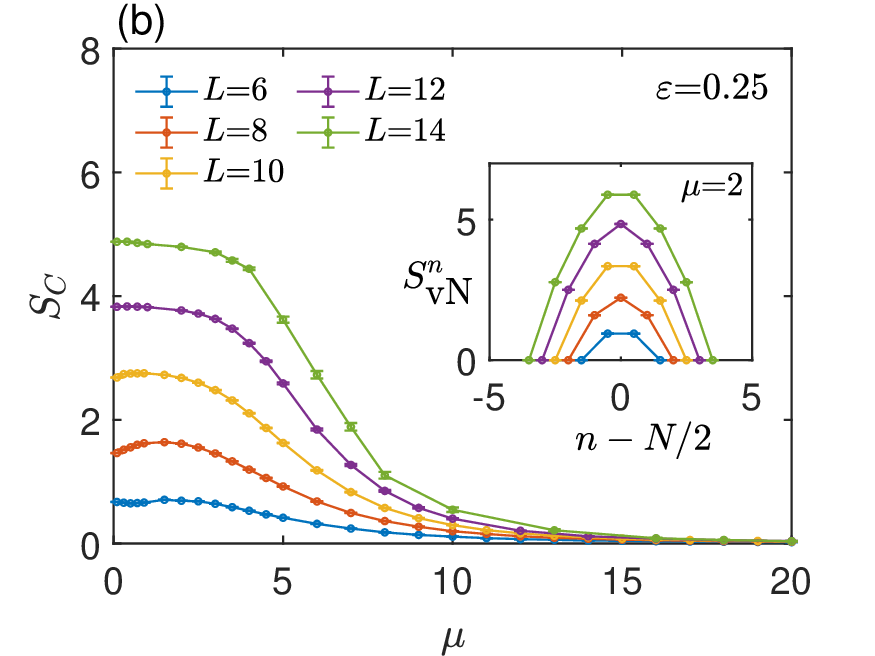}
		\caption{(a) Particle number entropy $S_N$ versus disorder intensity $\mu$ for different sizes $L$ at an energy density $\varepsilon=0.25$. (b) Results for configuration entropy $S_C$ at the same energy density. Companion results for $p_n$ and $S_{\textrm{vN}}^{n}$ with respect to differing sizes are shown in the insets of (a) and (b) for $\mu=2$.}
		\label{entropy}		
	\end{center}
\end{figure}

In Subsection~\ref{sec:subsect11}, based on the $U(1)$ symmetry, we decompose $S_{\textrm{vN}}$ into the particle number entropy $S_N$ (readily measurable) and the configuration entropy $S_C$ (not currently measurable). We first observe that the $S_C$ of the eigenstate nears zero in the localized phase, implying that $S_C$ may largely be phase transition irrelevant (or at least less relevant). Motivated by this, in Subsection~\ref{sec:subsect12}, we introduce a new quantity that quantifies the deviation of $S_N$ from the ideal thermal distribution. Subsequent finite-size analysis shows that it shares the same phase transition point with $S_{\textrm{vN}}$ but bears a greater critical exponent, indicating that it is $S_N$ that primarily drives the phase transition. Most importantly, this deviation of $S_N$ can be preliminarily measured in the laboratory, as has already been demonstrated by recent Harvard experiments \cite{35,leonard2023probing} (especially in boson systems). Admittedly, here we have also asserted the assumption that the energy-resolved eigenstate measurements we consider can be qualitatively executed through the careful combinations of disorder realizations and initial state preparations for using the dynamics, as was successfully implemented in the recent programmable-superconducting-processor-based experiment on MBL \cite{70}. Recent debates about whether the MBL transition is a true phase transition or a crossover \cite{vsuntajs2020quantum,morningstar2022avalanches} prompt us to seriously inspect such experimental measurability, and these considerations can contribute to the settling of these ongoing controversies (see comments at the end of this subsection). Finally, in Subsection~\ref{sec:subsect13}, to gain an overview of the eigenstate-transition scenario, we sketch the finite-system dynamical phase diagram obtained from the large-scale numerical simulations.

\subsection{Entanglement Entropy and Particle Number Fluctuation}
\label{sec:subsect11}

In Appendix~\ref{sec:append_Decomp}, we detail the decomposition of the total entanglement entropy $S_{\textrm{vN}}$ into the particle number entropy $S_N$ and the configuration entropy $S_C$ when the total number of particles $N$ in the system is conserved and discuss how to calculate the configuration entropy in an efficient way.

\begin{figure}[t]
	\begin{center}
		\includegraphics[scale=0.6]{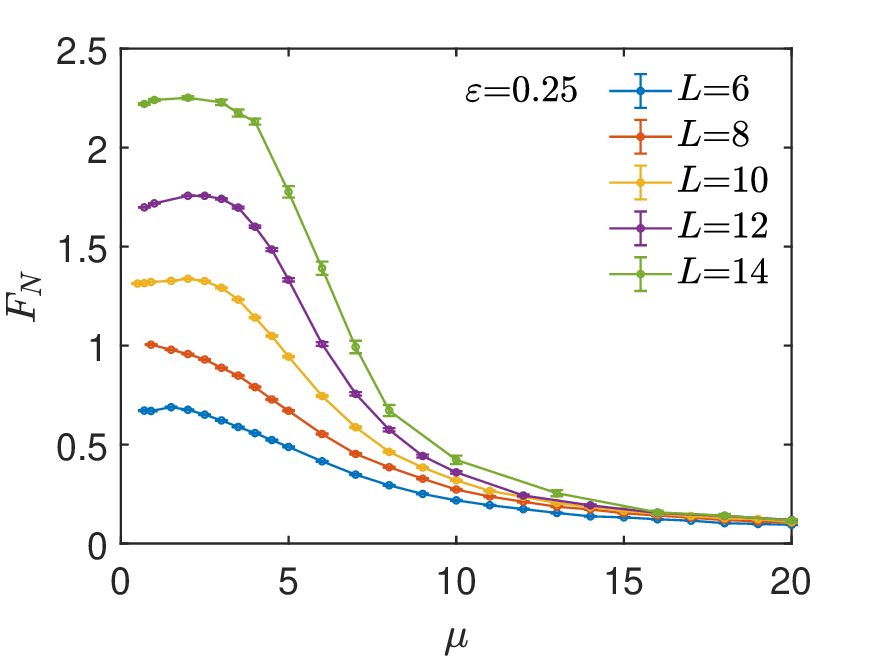}
		\caption{Half-chain particle number fluctuation $F_N$ versus $\mu$ at an energy density $\varepsilon=0.25$ for various lengths $L$.}
		\label{ParticleNumberFluctation}		
	\end{center}
\end{figure}

Now we mainly focus on the obtained numerical results for these two quantities, the particle number and configuration entropies. Figure~\ref{entropy}(a) shows the size dependence of $S_N$. For each length $L$, $S_N$ decreases monotonically with the increasing disorder strength $\mu$. Likewise, for $S_C$, we can see from Fig.~\ref{entropy}(b) that it lessens as disorder enhances, and approaches zero once the disorder strength $\mu$ becomes significant. This is because if disorder is strong enough, the system is expected to be localized in several particular configurations. The configuration entropy due to the superposition of these configurations thereby tends to near zero. Under the circumstance of larger $\mu$, the particle number entropy $S_N$ appears to dominate the evolution of the total entropy. Besides, both entropies exhibit a volume law at weak disorder and an area law at strong disorder, consistent with the general trend of the total entropy found in previous MBL studies \cite{11,13}.

As pointed out in Appendix~\ref{sec:append_Decomp}, $S_N$ and $S_C$ can be expressed in terms of $p_n$ and $S_{\textrm{vN}}^{n}$. Here, $n$ denotes the number of particles in the left half chain, marking the chunks of the reduced density matrix which we call channels. The value of $n$ ranges from $0$ to $N$ (there are a total of $N+1$ options for $n$). In the inset of Fig.~\ref{entropy}(a), we give the distribution of $p_n$ as a function of $n-N/2$ with respect to different chain lengths for a small disorder strength $\mu=2$. It can be seen that since $p_n$ needs to be normalized and there is no difference between tracing out the left or the right portion of the system at ETH, it is a symmetric distribution of the Gaussian lineshape. In the inset of Fig.~\ref{entropy}(b), the results of $S_{\textrm{vN}}^{n}$ are given, again symmetric. Overall, as the chain length increases, the dimension of the channel grows, so does the corresponding entropy.

$S_N$ arises from the superposition of states with varying particle numbers in half-chain subsystems, generated by particle motion across the boundary of the bipartite ($A$ and $B$). This fluctuation and its measurability can both be understood in terms of the particle number variances for part $A$, which is typically defined as follows,
\begin{eqnarray}
	\label{eq_FN}
	F_N&=&\left\{\langle\hat N_A^2\rangle-\langle\hat N_A\rangle^2\right\}_{\textrm{av}}\label{QEFN}\\
	&=&-\left\{\sum_{i<\frac{L}{2}}{\sum_{j>\frac{L}{2}}{\left[\langle\hat n_i\hat n_j\rangle-\langle\hat n_i\rangle\langle\hat n_j\rangle\right]}}\right\}_{\textrm{av}},\nonumber
\end{eqnarray}
where $\hat N_A$ is the total particle number operator for part $A$. $F_N$ accounts for the correlation between the two-particle densities on the two sites located inside portions $A$ and $B$, respectively, reflecting the particle-number-fluctuation induced entanglement between the bipartite.

Figure~\ref{ParticleNumberFluctation} shows the results of $F_N$ at $\varepsilon=0.25$. One can see that it indeed behaves analogously to $S_N$. This implies that it also has a volume law for small $\mu$ in the ETH phase and tends toward the area law at large $\mu$ for the MBL phase. Analyzing the last expression of (\ref{eq_FN}) when subject to the sufficiently large $\mu$, one may expect that the correlations would become exponentially small with increasing $|i-j|$. It turns out that in the summation of (\ref{eq_FN}), those terms with small $|i-j|$ do dominate, i.e., only the contributions of those correlations at the interface between parts $A$ and $B$ remain finite for the MBL phase, thereby leading to an area law for $F_N$. This also reinforces the experimental observation on the area-law scaling of $S_N$ in the MBL phase as was reported in the Harvard experiment \cite{35}.

\subsection{Number Entropy Dominates the MBL Transition}
\label{sec:subsect12}

As seen in the previous subsection, in the localized phase, the $S_C$ of the eigenstate nears zero and consequently $S_N$ dominates the total entropy. This implies that $S_C$ may be much less relevant to the phase transition and the transition is chiefly governed by $S_N$. Next, we will show numerical evidence to substantiate this claim.

\begin{figure}[tb]
	\begin{center}
		\includegraphics[scale=0.6]{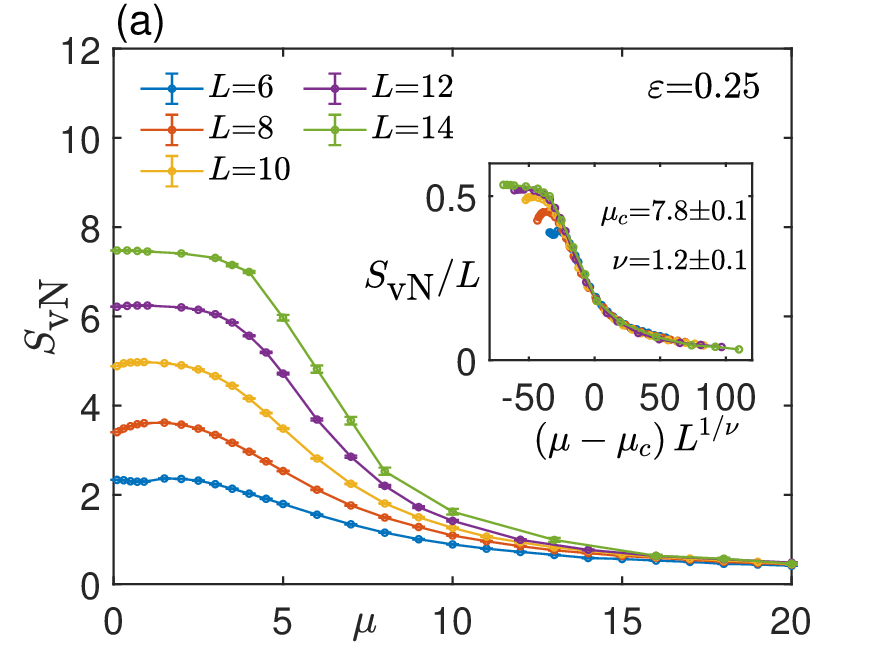}
		\includegraphics[scale=0.6]{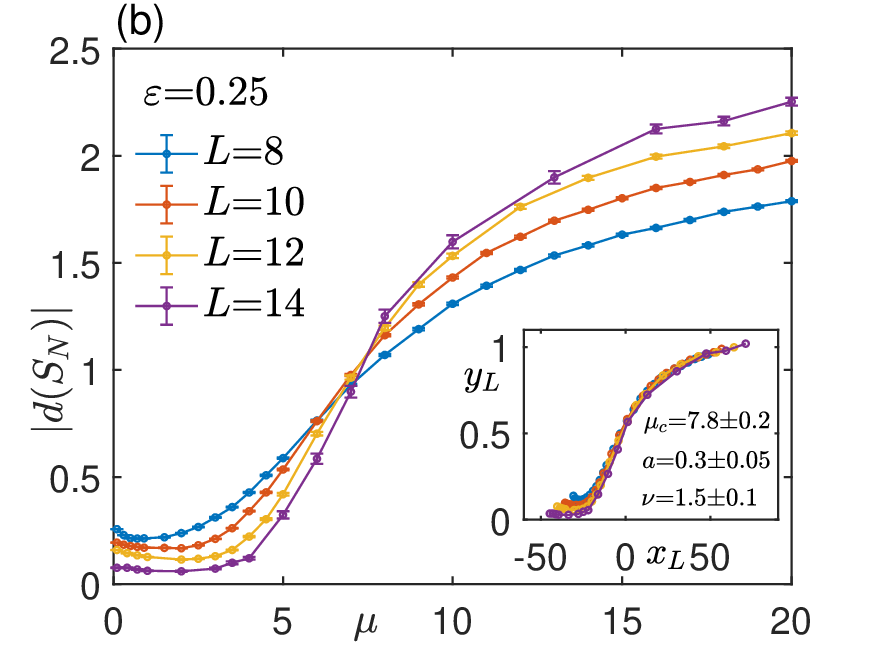}
		\caption{(a) Variation of the total entropy $S_{\textrm{vN}}$ with the disorder intensity for different sizes; the inset gives the result of its data collapse, yielding the estimate of the phase transition point at $\mu_c\approx7.8$. (b) is the corresponding result of $|d(S_{N})|$ for different chain lengths when disorder increases; its data collapse is given also in the inset, showing the same phase transition point at around $\mu_c\approx7.8$. The energy density here is always fixed to be $\varepsilon=0.25$.}
		\label{scaling}		
	\end{center}
\end{figure}

Firstly, we calculate the total von Neumann entropy $S_{\textrm{vN}}$ for a given $\varepsilon=0.25$ using the various sizes $L$, which is expected to capture the finer signal of the transition point according to the general scaling theory of the critical phenomena \cite{68,66}. As per the finite-size scaling theory, the phase transition point and the critical exponent thereof can be obtained when all data curves are collapsed into a single line describable by the following scaling form ansatz \cite{13,86},
\begin{eqnarray}
	Q(L,\mu)=g(L)f\bm{(}(\mu-\mu_c)L^{1/{\nu}}\bm{)},
\end{eqnarray}
which gives rise to a collapsed scaling relation of $y_L=Q( L,\mu)/g(L)$ versus $x_L=(\mu-\mu_c)L^{1/{\nu}}$ and the critical value $\mu_c$ as well as the critical exponent $\nu$ for the transition. $g(L)$ is generally a polynomial of $L$. Here, we set $g(L)=L$. In Fig.~\ref{scaling}(a), we show a size dependence of $S_{\textrm{vN}}$ at $\varepsilon =0.25$ for $L=6,8,10,12,14$. Using the scaling relation mentioned above, we plot $y_L=S_{\textrm{vN}}/L$ in the inset of Fig.~\ref{scaling}(a), where we can collapse the data into almost a single curve except for those points at the very low $\mu$. We are thus able to obtain a critical value $\mu_c\approx7.8$ and the associated critical exponent $\nu\approx1.2$, namely a phase transition point at $\varepsilon=0.25$.

Secondly, for an ideally thermalized system where the system has an equal probability to access various configurations, a classical probability can be obtained by dividing the dimension of each symmetry-channel (or symmetry-sector) block by the total dimension of the Hilbert space. For the present model, the total Hilbert space of a bosonic system with the chain length $L$ and the number of particles $N$ is $D_{N}^{L}=\frac{(L+N-1)!}{N!(L-1)!}$, and the dimension of the Hilbert subspace with particle number $n$ on the left half chain is $D_n=D_{n}^{L/2}D_{N-n}^{L/2}$, so $p_{n}^{\textrm{ideal}}=D_n/D_{N}^{L}$, and we can get $S_{N}^{\textrm{ideal}}=-\sum_{n=0}^N{p_{n}^{\textrm{ideal}}}\log_2 p_{n}^{\textrm{ideal}}$. To characterize the extent to which the actual probability distribution deviates from this ideal thermalization situation, we next define the following quantity so as to potentially locate the phase transition more accurately than the bare $S_N$, 
\begin{eqnarray}
	\label{eq_dpn}
	|d(S_N)|=|\{S_N\}_{\textrm{av}}-S_{N}^{\textrm{ideal}}|.
\end{eqnarray}
As shown in Fig.~\ref{scaling}(b), we take $g(L)=L^a$ for the case at hand, and one can see that the phase transition point thus obtained from $|d(S_N)|$ agrees with the phase transition point obtained from the total entropy $S_{\textrm{vN}}$ shown in Fig.~\ref{scaling}(a). 

More encouragingly, the critical exponent obtained from $|d(S_N)|$, $\nu\approx1.5$, is closer to satisfy the Harris bound $(\nu \geqslant2)$ \cite{65} compared to the critical exponent $\nu\approx1.2$ obtained from $S_{\textrm{vN}}$. Previous numeric evaluations of various physical quantities have yielded critical exponents all around $\nu\approx1.0$ \cite{88}, similar to that obtained from $S_{\textrm{vN}}$. This suggests that the inclusion of $S_C$ may weaken the critical phenomena in general and enhance the finite-size effects in particular. This could partially lead to the pronounced drifts in the finite-size calculations, making it challenging to distinguish between a phase transition and a crossover. Furthermore, it is noteworthy that we define the deviation $|d(S_N)|$ using $S_N$, and $S_N$ itself is computed by decomposing the total entropy as per the $U(1)$ symmetry of the system. It is this symmetry-based decomposition that allows us to separate out the phase-transition-less-relevant quantity $S_C$. This indicates that symmetry plays a crucial role in the MBL phase transition. Finally, the deviation $|d(S_N)|$ we defined can be accessed directly through the $S_N$ measurements, and recent experiments (especially in boson systems \cite{35,leonard2023probing}) have shown that $S_N$ might be available.

Currently, there has been substantial literature debating the existence of the MBL phase and whether the MBL transition is really a phase transition or merely a crossover \cite{vsuntajs2020quantum,vsuntajs2020ergodicity,morningstar2022avalanches}.
We do not assess their conclusions. However, based on the above analyses, the devised physical quantities in these works may have the following limitations. 
\begin{itemize}
\item[(1)] The models they studied generally have the $U(1)$ symmetry, but the computed physical quantities do not incorporate this symmetry thoroughly. Even worse is the case that the computational scheme they used may explicitly break those important internal symmetries of the system, such as the open-system approach used in \cite{morningstar2022avalanches} which spoils the $U(1)$ symmetry of the Heisenberg chain. This lack of the symmetry embedding, particularly in a way parallel to the entropy decomposition already implemented here in Appendix~\ref{sec:append_Decomp}, may result in the dangerous inclusion of the phase-transition-irrelevant components or contributions, which probably distorts the revealing of the underlying critical phenomena.
\item[(2)] Most quantities designed might not be experimentally measurable.
\item[(3)] Various transport calculations require to involve the information of the whole energy spectrum. As will be illustrated shortly, the higher- and lower-energy behaviors of the generic bosonic systems can be notably distinct. This means that these regimes need to be considered and examined carefully and separately. For instance, when calculating the transport properties of a bosonic chain at infinite temperature, blindly mixing all eigenstates equally may pose a problem.
\end{itemize}
Following the recent theoretical framework proposed by \cite{morningstar2022avalanches}, the observed MBL phase is likely akin to what they call the finite-size MBL regime, with the phase transition point corresponding to the landmark $\mathcal{L}_r$ within the finite-size accuracy. In essence, while debates over the MBL phase and transition persist, we provide an experimentally practical quantity to aid the investigation of these critical issues.

\subsection{Finite-System Dynamical Phase Diagram}
\label{sec:subsect13}

\begin{figure}[t]
	\begin{center}
		\includegraphics[scale=0.6]{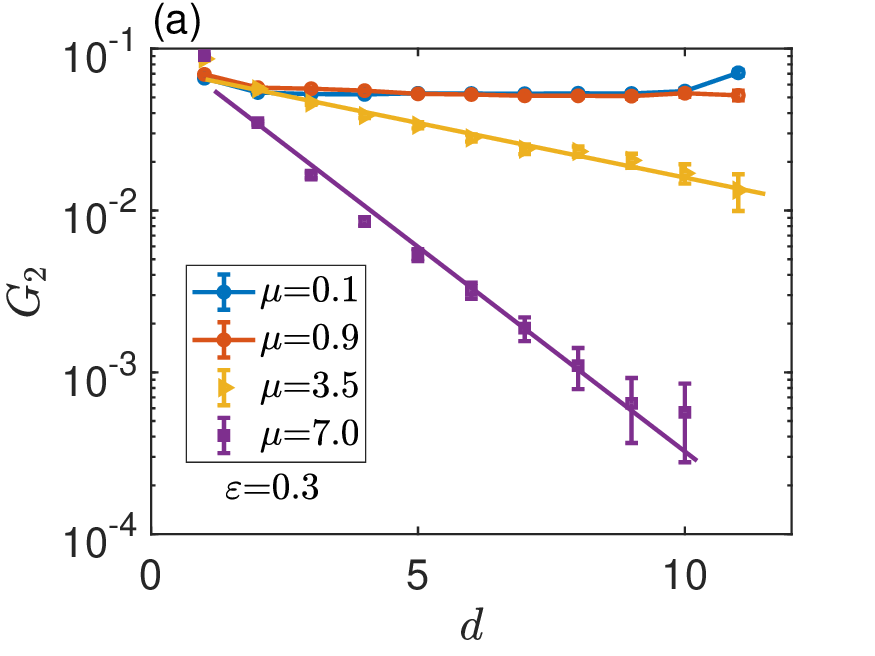}
		\includegraphics[scale=0.6]{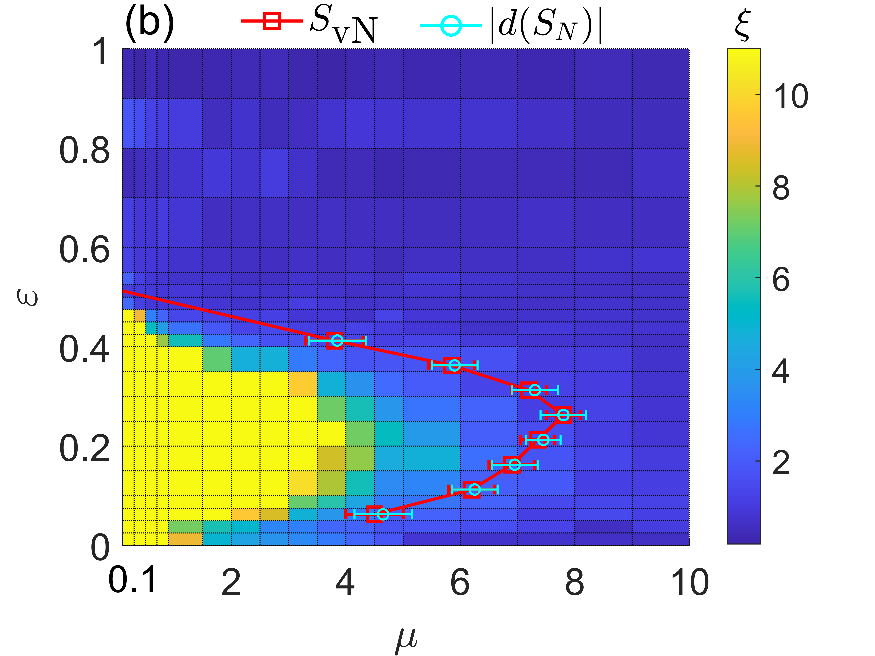}
		\caption{(a) Numerical results on the two-body density-density correlation function $G_2$ for $L=12$ from the ED calculations. $G_2$ versus $d$ shows an exponential decay for $\mu=3.5$ and $7.0$ at $\varepsilon=0.3$, from which the localization lengths might be extracted. (b) Phase diagram on the plane of energy density $\varepsilon$ and disorder strength $\mu$ is constructed based on the color contour of the localization length $\xi$ and consists of the ETH and MBL regimes, separated by a line derived from the scaling analyses of the total entropy $S_{\textrm{vN}}$ and $|d(S_N)|$.}
		\label{TwoBodyCorrelation}		
	\end{center}
\end{figure}

\begin{figure}[t]	
	\begin{center}
		\includegraphics[scale=0.6]{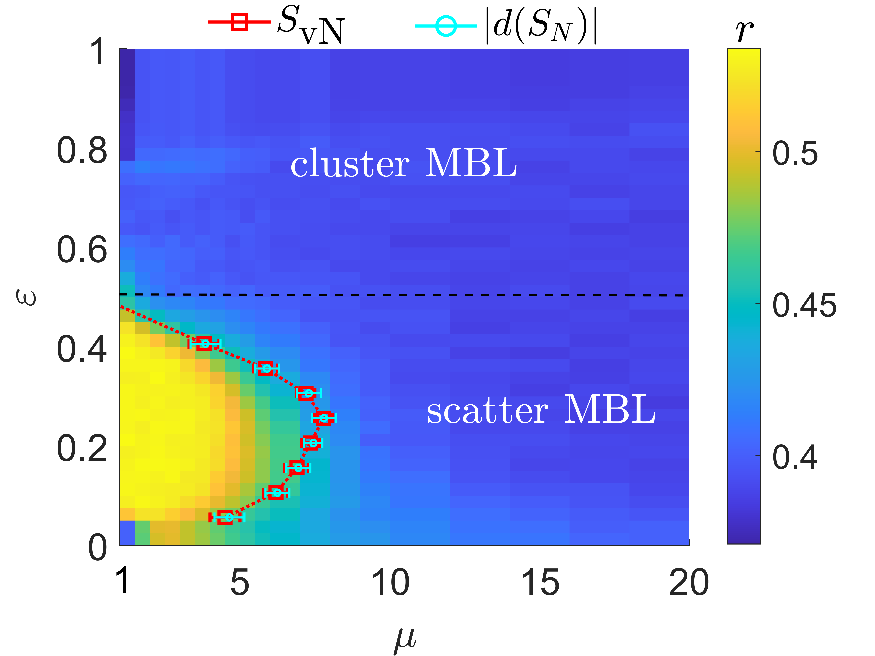}
		\caption{Landscape of the phase diagram from the adjacent gap ratio $r$, using a dBH chain of $L=12$. The right color bar labels $r$ approximately between the GOE limit $r_{\textrm{GOE}}\approx0.536$ and the Poisson limit $r_{\textrm{Poi.}}\approx0.386$. A blue square in the bottom-left corner is an artifact that may occur from an insufficient number of eigenstates, particularly in such an area where both $\varepsilon$ and $\mu$ are small, i.e., essentially originating from the finite-size effect. In addition, schematics of the different MBL regions are indicated: scatter MBL and cluster MBL. The red (from $S_{\textrm{vN}}$) and cyan (from $|d(S_N)|$) squared symbols with the horizontal error bars for $\mu$ mark the position estimates of the phase boundaries at $\varepsilon=0.05,0.1,0.15,0.2,0.25,0.3,0.35,0.4$.}
		\label{gapratio}		
	\end{center}	
\end{figure}

Based on the scaling analyses discussed in the previous subsection, we analogously extracted the phase transition points for a spectrum of other energy densities, ultimately producing the eigenstate phase transition diagram for the dBH model, as shown by Figs.~\ref{TwoBodyCorrelation}(b) and \ref{gapratio}. It can be seen that $S_{\textrm{vN}}$ and $|d(S_N)|$ yield nearly identical estimates for the phase boundaries.

Inspired by the experimental pertinence, we may use the measurable localization length to illustrate that the system's thermalization states predominantly reside in the low-energy-density region. Firstly, the two-body density-density correlation function $G_2(d)$ is introduced to measure the correlation of the particles located at the two sites separated by a distance $d$ in an eigenstate with the eigenenergy closest to a given energy density $\varepsilon$. Specifically, we borrow the definition from the Harvard experiment \cite{35}
\begin{eqnarray}
G_2(d)=-\left\{\frac{\sum_{i+d\leqslant L}{[\langle\hat n_i\hat n_{i+d}\rangle-\langle\hat n_i\rangle\langle\hat n_{i+d}\rangle]}}{N_{i+d\leqslant L}}\right\}_{\textrm{av}}
\end{eqnarray}
where $N_{i+d\leqslant L}$ indicates the number of pairs in the summation over $i$ and the given $d$, and $\left\{ ... \right\}_{\textrm{av}}$ denotes an average over the disorder samples with a given set of $\mu$ and $\varepsilon$. Since the total particle number is conserved for the present dBH model, one generally has $G_2(d)\geqslant 0$. Further, one expects that $G_2(d)$ decreases exponentially with $d$ when the disorder strength is large enough. The localization length can thereby be extracted from this exponentially decaying behavior of $G_2(d)$, alternatively providing a distinguishable signal between the ETH and MBL phases.

In Fig.~\ref{TwoBodyCorrelation}(a), the two-body density-density correlation function at $\varepsilon =0.3$ is shown for the four disorder strengths. In the region of small disorder strength, the correlation function decays very slowly with the distance $d$. It is nearly a constant, indicating that all the system's parts are highly entangled and the many-body eigenstates are well thermalized. As the disorder strength increases, an exponential decay behavior becomes sharper and sharper so that it can be fitted by a formula of the form $G_2(d) =Ae^{-d/\xi}$ and the localization length $\xi$ can be extracted thereof. (Note, in comparison, the localization length cannot be gained via fitting to an exponential form when the disorder is weak and the energy density is small.) Accordingly, the emergence and the change of the localization length tunable by $\mu$ reveal an eigenstate phase transition from the ETH state to the MBL state.

As illustrated by Fig.~\ref{TwoBodyCorrelation}(b), by exploiting the extracted localization lengths, for the 1D dBH model with $U/J=3$ and the half-filling, we find an ETH phase in the bottom-left corner (yellow color) and the MBL phase (blue color) in the rest of the $\mu$-$\varepsilon$ parameter plane. As already stressed, the thermalized states predominantly distribute within the low-energy, weakly disordered region of the phase diagram.

Moreover, as an aside, when comparing the phase transition boundary to the color map of the localization length, one notices that the localization lengths are generally much greater than several lattice spacings in the ETH phase. Nonetheless, we would like to remind the reader the procedure that for a thin area of the very weak disorder strengths in the ETH phase, the values of these large localization lengths that cannot be determined by using the exponential fit of $G_2$ have all been truncated to the $L-1$ lattice spacings, the maximal possible value of $d$, so that one can still represent the ETH phase in Fig.~\ref{TwoBodyCorrelation}(b) in a coherent fashion.

To further confirm the construct of the phase diagram, it is instructive to study the complementary energy-resolved adjacent gap ratio defined by \cite{76,77,orell2019probing}
\begin{equation}
	r_n=\frac{\textrm{min}[\delta E_n,\delta E_{n+1}]}{\textrm{max}[\delta E_n,\delta E_{n+1}]}
\end{equation}
where $\delta E_n=E_n-E_{n-1}$ is the difference between the two successive eigenenergies arranged in an ascending order. For each disorder sample, we calculate the $16$ adjacent $r_n$ to obtain the averaged value $\bar{r}$ subject to a given energy density $\varepsilon$, and $r=\{\bar{r}\} _{\textrm{av}}$ denotes the average over the disorder samples. The ratio $r$ reflects how the disorder strength $\mu$ determines the nature of the level statistics at a specific energy density $\varepsilon$. Figure~\ref{gapratio} shows the landscape of $r$ versus the energy density $\varepsilon$ and the disorder strength $\mu$ with $L=12$. One can see that the ETH state is located in the yellow region with $r$ approaching the Gaussian orthogonal ensemble (GOE) limit $r_{\textrm{GOE}}\approx 0.536$. By contrast, the MBL state, represented in blue, tends to reach the Poisson limit $r_{\textrm{Poi.}}\approx 0.386$ \cite{81}. The results of the phase boundaries, obtained from finite-size analyses of $S_{\textrm{vN}}$ and $|d(S_N)|$, are also overlaid in this phase diagram for comparison. From Fig.~\ref{gapratio}, it is apparent that there appears to be an intermediate region between the yellow-region boundary and the phase-transition boundary. In Appendix~\ref{sec:append_Comp}, we explore this low-energy and medium-$\mu$ region further from the perspective of the local compressibility distribution.

In the region where the energy density is less than $0.5$, the overall phase transition resembles that in the full spin or fermion models \cite{10,11,12}. It is anticipated that when the energy density is relatively low, it is not easy to form the high-energy situation where multiple particles occupy the same position. While, in the high-energy-density region, particles can and do form clusters due to interaction and bosonic statistics, yet possibly with very different behaviors between the limits of small and large disorder strengths.

As shown by Figs.~\ref{TwoBodyCorrelation}(b) and \ref{gapratio}, there is basically no significant change in the color of the localized part of the diagram. But we are now equipped to elucidate the properties and distinctions of these phase regions in some detail by scrutinizing the eigenstate properties and the quench dynamics in the high- and low-energy portions of the phase diagram. These peculiarities of the dBH model, as will be described next in Section~\ref{sec:sect4}, exemplify the intricate interplay of interaction, quantum statistics, and disorder, and give the schematic delineation of the phase regions in Fig.~\ref{gapratio} as per the broad classification of scatter and cluster MBLs.

\section{\label{sec:level1}Many-Body Localized Eigenstates}
\label{sec:sect4}

This section focuses on the properties of MBL eigenstates in the dBH chain. Our investigations reveal the following findings. (i) In Subsection~\ref{sec:subsect22}, we point out that the low-energy-density portion of the eigenspectrum of the dBH model closely resembles the entire eigenspectrum of the fermionic (or spin) models \cite{10,11,12}, featured by a transition from the thermalized to localized states. In view of the fact that the low-energy mobile identities are scatter particles, we name the resulting localized state the scatter MBL phase. In Subsection~\ref{sec:subsect22}, we analyze the eigenstate particle number distribution for scatter MBL and summarize the associated quench dynamics characteristics. (ii) Owing to the bosonic statistics, the high-energy-density portion of the eigenspectrum allows the formation of the distinct boson clusters in a random potential background. This situation is unique to the dBH-type model, and we name the resulting localized state the cluster MBL phase. After examining the eigenstate particle number distribution and the quench dynamics features, we show that the cluster MBL may differ in significant aspects from the scatter MBL. (iii) Regarding the localized eigenstate properties, the traditional measures might not be directly sensitive to the differences between the high- and low-energy-density sections of the eigenspectrum. In Subsection~\ref{sec:subsect24}, by exploring the structure of the symmetry-resolved entanglement entropy $S_{\textrm{vN}}^{n}$, we introduce another measure to help discriminate between the eigenstate phases emerging at the respective low and high energies.

\subsection{Scatter MBL in the Low-Energy-Density Spectrum}
\label{sec:subsect22}

First, we define the following two states that will be used below. Note, the subscripts in (\ref{lstate}) and (\ref{pstate}) label the site indices.
\begin{itemize}
\item[(1)] A line-shape initial product state, called the $l$-state, consists of one boson on each site of the left half chain. Pictorially, it is given by 
\begin{equation}
|l\rangle=|1_1,1_2,\ldots,1_{\frac{L}{2}},0_{\frac{L}{2}+1},\ldots,0_L\rangle.
\label{lstate}
\end{equation}
\item[(2)] A point-shape initial product state, called the $p$-state, is formed by putting all $N$ bosons onto the leftmost site. It is given by
\begin{equation}
|p\rangle=|N_1,0_2,\ldots,0_L\rangle.
\label{pstate}
\end{equation}
\end{itemize}
Their corresponding energy densities are shown in Appendix~\ref{sec:append_Aver}. From there, it is clear that as disorder increases, the $l$-state always stays within the low-energy-density region, while the $p$-state, in comparison, always stays within the high-energy-density region.

\begin{figure}[tb]
	\begin{center}
		\includegraphics[scale=0.98]{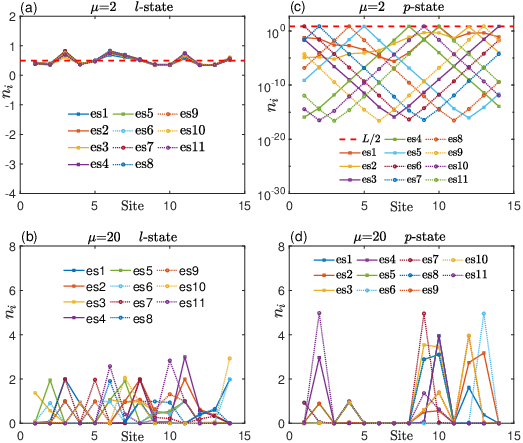}
		\caption{Particle number distributions of the 11 eigenstates, 5 above and 5 below the eigenstate closest in energy to the given initial state. (a) corresponds to the target energy $E = \left\langle l|H|l \right\rangle_{\mu=2}$ realized in the region of low energy density and small disorder strength. (b) corresponds to $E = \left\langle l|H|l \right\rangle_{\mu=20}$ within the low-energy and strong-disorder region. (c) gives the result of $E = \left\langle p|H|p \right\rangle_{\mu=2}$ for the high energy density and weak disorder. (d) targets $E = \left\langle p|H|p \right\rangle_{\mu=20}$ upon the high energy density and large disorder strength. The chain size here is fixed to be $L=14$.}
		\label{par_dis}		
	\end{center}
\end{figure}

Selecting a random sample, we first compute the energy of the $l$-state as per $E = \left\langle l|H|l \right\rangle_{\mu=2}$. Next, we identify the 11 nearest eigenstates (5 above and 5 below the target eigenstate whose eigenenergy is closest to $E$ given above). Subsequently, we calculate the particle number distributions for these eigenstates at each lattice site. In Fig.~\ref{par_dis}(a), the particle numbers at weak disorder $(\mu=2)$ are predominantly distributed around $0.5$. This is reasonable because the system is in a thermalized state whose particle numbers are distributed uniformly across the lattice sites. Consider that we are dealing with a half-filled system, then the particle number should indeed be at about $0.5$. In \cite{84}, we observe that for the weak disorder at $\mu=2$, starting from the $l$-state, $S_N$ grows with time as a function of $\ln t$, and its steady-state value grows with size as per $\ln L$. On the other hand, the companion $S_C$ grows with time linearly, and exhibits a volume-law scaling behavior at the long-time steady-state limit.

We repeat the same calculation for the $l$-state but under the condition of strong disorder, i.e., $\mu=20$. As shown by Fig.~\ref{par_dis}(b), in this scenario, some lattice sites exhibit particle clustering, while others remain nearly empty, indicating the particle localization within the chain. It is worth noticing that the maximum particle occupancy per site in Fig.~\ref{par_dis}(b) remains low, and the situation where the majority of the particles occupy a single lattice site does not occur. We call this low-energy MBL phenomenon the scatter MBL. In \cite{84}, we show that for this strong-disorder condition at $\mu=20$, if starting from the $l$-state, the growth of $S_N$ slows down to $\ln\!\ln t$, and the scaling of its steady-state value obeys the area law. Correspondingly, the companion $S_C$ in this situation grows with time as per a function of $\ln t$, and its steady-state value fulfills a volume law. 

Incidentally, with the channel-resolved analysis, we show that this slow $\ln\ln t$ growth of $S_N$ is mainly owing to the localization phenomenon and its final steady-state scaling remains to be an area law \cite{79}. These observations are in contradiction with the prediction of the eventual thermalization for the similar situation in the disordered spin-chain model as was claimed in recent literature \cite{80}.

In the recent seminal experimental work on the dBH model \cite{35}, it was observed that within the MBL phase, $S_N$ follows an area-law scaling behavior, while concurrently it was implicit that $S_C$ follows a volume-law scaling behavior ($S_C$ cannot be measured directly, so a closely related correlator was measured experimentally as an alternative). This suggests that the low-energy scatter MBL behavior has been experimentally detectable, but to our knowledge, the subsequent discussion regarding the high-energy cluster MBL behaviors has never been reported before.

\subsection{Cluster MBL in the High-Energy-Density Spectrum}
\label{sec:subsect23}

We do the same calculation by switching the initial state from the low-energy $l$-state to the high-energy $p$-state. For weak disorder $(\mu=2)$, as depicted in Fig.~\ref{par_dis}(c), we observe that the 11 nearest eigenstates at the target energy $E=\left\langle p|H|p \right\rangle_{\mu=2}$ have a common feature in their particle number distributions. Concretely, almost all the particles are concentrated to occupy a single lattice site, while the particle numbers at other sites decay exponentially. This signifies a strong degree of localization within the system; we call it the hard-type cluster MBL. This is because, in the high-energy-density regime, Bose statistics allows for the formation of the multi-particle clusters. Without disorder, these clusters can superpose to form the cat like states, but disorder breaks the translational symmetry, thus obstructing their coherent superposition and consequently giving rise to the localized, isolated clusters, ultimately resulting in the robust cluster MBL.

In \cite{84}, we find that, for the situation of low disorder $(\mu=2)$, starting from the $p$-state, both $S_N$ and $S_C$ exhibit almost no growth over time, with the steady-state values following the area law. Loosely speaking, these behaviors resemble the Anderson localization in some aspects.

Similar calculations upon the $p$-state are performed under the strong-disorder condition $(\mu=20)$ as well. It can be seen from Fig.~\ref{par_dis}(d) that, at this circumstance, the system also exhibits the distinct clustering behavior. But it is no longer as extreme as in Fig.~\ref{par_dis}(c), where nearly all particles are localized onto a single point, nor is it close to Fig.~\ref{par_dis}(b), where particles are localized to only reach the relatively low local concentrations. Therefore, we refer to this situation as the soft-type cluster MBL.

In \cite{84}, we demonstrate that for this strong disorder at $\mu=20$, starting from the $p$-state, $S_N$ exhibits the diminished time-dependent growth, with the steady-state values following an area law. By contrast, the companion $S_C$ keeps increasing slowly over time in a tentative $\ln\!\ln t$ format, but its steady-state values seem to still obey the area law.

\subsection{Fine Structure of $S_{\textrm{vN}}$}
\label{sec:subsect24}

Curiously, is there a single quantity that can both reveal the thermalized-state distribution and differentiate between the behaviors within the low- and high-energy-density sections of the phase diagram? Through exploiting the channel reflection symmetry of the set $\{S_{\textrm{vN}}^{n}\}$, we introduce a new physical quantity that effectively addresses this question. Complementarily, this might provide a different means to probe the phase transition zone from the pure quantum entanglement $\{S_{\textrm{vN}}^{n}\}$ via its intrinsic structure.

\begin{figure}[tb]
	\begin{center}
		\includegraphics[scale=0.6]{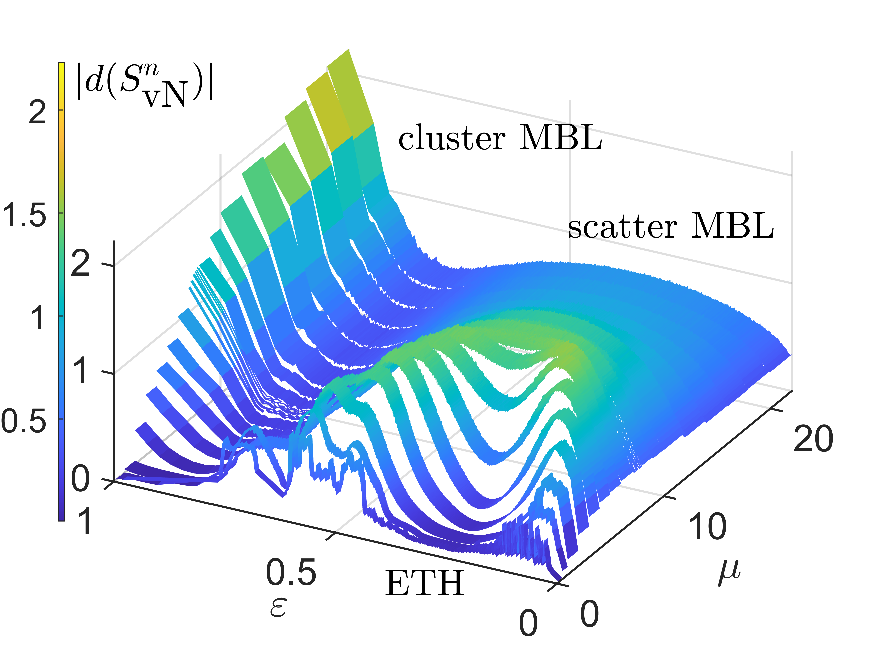}
		\caption{3D plot of $|d(S_{\textrm{vN}}^{n})|$ at $L=12$, where the phase transition boundary and the three different phases can be perceived.}
		\label{dsvnn_3d}		
	\end{center}
\end{figure}

As claimed in Appendix~\ref{sec:append_Decomp}, for the symmetry-resolved entropy $\{S_{\textrm{vN}}^{n}\}$, the relation $S_{\textrm{vN}}^{n_A=n}=S_{\textrm{vN}}^{n_B=N-n}$ always holds. In the absence of disorder, the preserved spatial reflection symmetry guarantees the general satisfaction of $S_{\textrm{vN}}^{n_A=n}=S_{\textrm{vN}}^{ n_A=N-n}$, which generates a channel reflection symmetry of $\{S_{\textrm{vN}}^{n}\}$, as already illustrated in the inset of Fig.~\ref{entropy}(b) above. It turns out that we can determine the degree of deviation from this symmetry by the summed difference of the corresponding pairs of channels, which is equivalent to determining the degree of deviation from the ideal thermalization. Thus, we try to define the pertinent physical quantity as follows,
\begin{eqnarray}
\label{eq_dsvnn}
|d(S_{\textrm{vN}}^{n})|=\left\{\sum_{n=1}^M{|S_{\textrm{vN}}^{n}-S_{\textrm{vN}}^{N-n}|}\right\} _{\textrm{av}},
\end{eqnarray}
where
\begin{eqnarray}
\label{eq_dsvnn_M}
M=\left\{ \begin{array}{l}
(N-2)/2\ \ \ \mbox{for}\ N=\mbox{even} \\[0.5em]
(N-1)/2\ \ \ \mbox{for}\ N=\mbox{odd} \\
\end{array} \right..
\end{eqnarray}
Since $S_{\textrm{vN}}^{n_A=0}=S_{\textrm{vN}}^{n_A=N}=0$, the summation starts from $n=1$. In the ETH region, the left and right parts are approximately symmetric, giving rise to a small value of $|d(S_{\textrm{vN}}^{n})|$. In the scatter MBL region, although the relative difference is nonzero, the value of $|d(S_{\textrm{vN}}^{n})|$ remains small because of the absolute smallness of the values of $\{S_{\textrm{vN}}^{n}\}$ themselves. Taken together, these trends produce a peak in the middle, corresponding to the critical regime of the underlying phase transition.

Further, from Fig.~\ref{dsvnn_3d}, we can see that the ETH region corresponds to the valley in the 3D figure of $|d(S_{\textrm{vN}}^{n})|$. The closer to the bottom of the valley, the lower the degree of the channel-reflection-symmetry breaking is and the more thorough the thermalization of the system assumes. There are two peaks at the valley's boundaries, corresponding to the location of the energy-resolved phase transition. When the two peaks merge and disappear, $|d(S_{\textrm{vN}}^{n})|$ becomes flattened, signalling that the system begins to enter the scatter MBL region. Between the high- and low-energy-density sections, a distinctive dip effectively separates the cluster MBL at high energies and the scatter MBL at low energies. In this regard, $|d(S_{\textrm{vN}}^{n})|$ may possess the capacity to distinguish between these MBL phases and aptly identify the thermalized region at the low energies.

The above content indicates that the internal structure of the entanglement entropy effectively distinguishes between different phase regions, upon resort to the concept of symmetry decomposition. Once again, one can appreciate the importance of \lq\lq symmetry combined with entanglement'' in the study and elucidation of the MBL phases and transitions.

\section{\label{sec:level1}Summary and Discussions}
\label{sec:sect6}

This paper provides a comprehensive study on the eigenstate properties of the dBH model in one dimension. 

Firstly, we explore the transition from the thermalized state to the MBL state at low energies. The von Neumann entropy, $S_{\textrm{vN}}$, is often used to detect the MBL transition, even though it cannot be directly measured. Utilizing $U(1)$ symmetry, the von Neumann entropy is decomposed into the particle number entropy, $S_N$, and the configurational entropy, $S_C$. Notably, $S_C$ approaches zero within the localized phase. We take advantage of this trend and define $|d(S_N)|$ to quantify the deviation of $S_N$ from the ideal thermal distribution. Finite-size scaling analysis shows that $|d(S_N)|$ shares the same critical point with $S_{\textrm{vN}}$ but exhibits a larger critical exponent, suggesting that the transition is primarily set by $S_N$ and its fluctuations. It is worth noting that $|d(S_N)|$ might be measured via protocols already used by recent experiments, particularly in the case of bosonic systems. Recent debates about whether MBL transitions are true phase transitions or crossovers prompt us to seek such an experimentally measurable quantity that can contribute to addressing these concerns.

Secondly, the paper contrasts the distinctive behaviors occurring in the high- and low-energy-density regimes of the phase diagram for the dBH model. We find that: (i) The thermalizing states predominantly occupy the low-energy spectrum, as evidenced by energy density distribution, localization lengths, and gap ratios. (ii) The low-energy spectrum closely resembles the fermionic (or spin) models, where the transition from the thermalizing state to the scatter MBL phase is visible. We analyze the characteristics of these low-energy localized eigenstates, and their dynamics are also summarized. (iii) Bose statistics leads to the formation of distinct clusters in the high-energy spectrum, resulting in the unique cluster MBL phase that significantly differs from the scattering MBL phase in low energies, as reflected through the eigenstate and dynamics properties. (iv) Traditional measures such as localization length and gap ratios cannot directly reveal the differences in the localization phenomena between the high- and low-energy sections. We define a new quantity, $|d(S_{\textrm{vN}}^{n})|$, from the internal structure of $S_{\textrm{vN}}$, which effectively captures the distinctions between phase regions in high and low energies.

The two quantities we defined based on the decomposition of entropy by symmetry, $|d(S_N)|$ and $|d(S_{\textrm{vN}}^{n})|$, prove to be useful in analyzing critical points and distinguishing between different phase regions. This shows that symmetry plays a key role in MBL phases and phase transitions. Currently, a grand portion of the MBL research has not given due attention to this aspect, and the physical quantities employed to analyze MBL phases and transitions have not properly embedded the symmetry information, particularly in those contentious issues.

As can be noticed, there needs a further study on whether the MBL regions are separated by phase transitions or crossovers. Moreover, the bimodal compressibility distribution structure seen in Appendix~\ref{sec:append_Comp} seems to predict the existence of a critical regime arising from the Griffiths effect \cite{26,69,66,67,73}. Different implementations of the randomized samples, e.g., quasiperiodic fields, may give rise to qualitatively different results \cite{63,64,65}. Although the numerical resource to precisely settle the localization problem for the dBH model is far beyond the current computational capacity, one might still imagine to resort to the available finite-size numerics to search for the emergent new physics when the filling factor and the Hubbard interaction are tuned within the wider parameter ranges.

\begin{acknowledgments}

We thank Zi Cai, Changfeng Chen, Shijie Hu, Haiqing Lin, Mingpu Qin, Jie Ren, and Wei Su for fruitful discussions. This work is supported by the Ministry of Science and Technology of China (Grant No. 2016YFA0300500) and the NSF of China (11974244). X.Q.W. also acknowledges additional support from the Shanghai talent program. C.C. acknowledges the support from the SJTU start-up fund and the sponsorship from Yangyang Development fund.

\end{acknowledgments}

\appendix

\section{\label{sec:level1}Decompose and Compute the Entanglement Entropy}
\label{sec:append_Decomp}

Consider that the system consists of two parts, $A$ and $B$. For convenience, we choose the system size $L$ to contain an even number of sites, and the sites for both parts $A$ and $B$ are equal, i.e., $L_A=L_B=L/2$ in the present work. For a given many-body wavefunction $|\psi(i_A,i_B)\rangle$, one may construct a reduced density matrix $\rho$ for part $A$ by tracing out those bases of part $B$, which can be written formally as $\rho(i_A,j_A)=\sum_{i_B}^{D_B}|\psi(i_A,i_B)\rangle\langle\psi(j_A,i_B)|$ where $D_B$ is the dimension of the Hilbert space for part $B$, and $i_A$ and $i_B$ are the indices to denote the bases of parts $A$ and $B$, respectively. The von Neumann entanglement entropy is thus defined for this system as $S_{\textrm{vN}}=-{\rm Tr}\rho{\log_2}\rho$, which is widely adapted to explore the natures of both the ETH and the MBL phases, caused by many-body interactions and disorder. One finds that $S_{\textrm{vN}}$ can be written as a sum of a particle number entropy $S_N$ and a configuration entropy $S_C$, which account for the entanglements of the particle number fluctuations and the correlations of particles between the two parts $A$ and $B$, respectively. Although $S_{\textrm{vN}}$ is not yet directly accessible in experiments, the recent Harvard experiment has successfully measured the particle number entropy $S_N$ and the configuration correlators that are somehow equivalent to the configuration entropy $S_C$, and found the area law of $S_N$ and the volume law of the configuration correlators in time evolving a state within the MBL phase \cite{35}.

For the Hamiltonian (\ref{dbhmodel}), since $[\hat H,\hat N]=0$ where $\hat N=\sum_{i}^{L}\hat n_i$, then $[\rho,N_A]=0$ where $N_A$ is the matrix representation of $\hat N_A=\sum_{i}^{L_A}\hat n_i$. This implies that the reduced density matrix is block diagonal, which can be written as a direct summation of the form $\rho=\oplus_{n=0}^{N}\rho^{(n)}$. Each $\rho^{(n)}$ has a dimension of $D^A_n$, representing the number of those bases having $n$ particles in the part $A$. It turns out that one can naturally explore the relative entropy for each given $n$, i.e., the particle number entropy, if one introduces the probability $p_n$ to sum up the amount of those states with $n$ particles, as given by $p_n={\rm Tr} \rho^{(n)}$. Subsequently, one may define a proper reduced density matrix $\rho_n=\rho^{(n)}/p_n$ for $n$, which obeys the compulsory normalization condition ${\rm Tr}\rho_{n}=1$. The reduced density is thus rewritten as $\rho=\oplus_{n=0}^{N}p_n\rho_n$.

In terms of $p_n$ and $\rho_n$, the von Neumann entropy can be expressed as
\begin{equation}
	S_{\textrm{vN}}=-\sum_{n=0}^{N}[p_n\rho_n]{\log_2}[p_n\rho_n]
\end{equation}
and can be further decoupled into a sum of the two parts $S_N$ and $S_C$ \cite{35}, which are given as follows,
\begin{eqnarray}
	S_N&=-&\sum_{n=0}^{N}p_n{\log_2}p_n,\\
	S_C&=-&\sum_{n=0}^{N}p_nS_{\textrm{vN}}^{n},
\end{eqnarray}
where $S_{\textrm{vN}}^{n}$ is called the symmetry-resolved entropy, and $S_{\textrm{vN}}^{n}=-\sum_{i=1}^{D_n}\rho_n(i,i){\log_2}\rho_n(i,i)$, where $\rho_n(i,i)$ is the normalized eigenvalue of $\rho^{(n)}$.

$S_{\textrm{vN}}^{n}$ is a partial configuration entropy that measures the entanglement between the two parts $A$ and $B$, resulting from the dynamics of particles' virtual superpositions and hoppings without changing $n$. Thus, each partial configuration entropy $S_{\textrm{vN}}^{n}$ for that $n$ carries a weight $p_n$, the sum of which forms the total configuration entropy $S_C$. $S_N$, on the other hand, stands for the entanglement of the two parts $A$ and $B$ associated to the particle number fluctuations encoded in the given $|\psi(A,B)\rangle$ under the constraint that the filling factor of the system is fixed.

Next, let us discuss an optimization issue. We found that it is practically significant to construct the bases optimally for both parts $A$ and $B$, which enables us to treat each $\rho^{(n)}$ directly so as to access larger-size systems in the most economic usage of the memory and the CPU resource. For this purpose, the Hilbert space for part $A$ is naturally assigned into the sequential subspaces, with an ascending order of the particle number from $0$ to $N$, i.e., the $(n+1)$-subspace contains $n$ particles with a dimension of $D_n^A$. The Hilbert space for part $B$ is also constructed similarly for consistency.

Based on the above construction of the bases, one can consider $|\psi(i_A,i_B)\rangle$ as a rank-$2$ tensor with indices $i_A$ and $i_B$ which denote the allocation site of each basis in the Hilbert spaces of the parts $A$ and $B$, respectively, with dimensions $D_A=\sum_{n=0}^{N}D^A_n$ and $D_B=\sum_{n=0}^{N}D^B_n$. Consequently, this tensor can contain rectangular sub-tensors $|\varphi(i_A,i_B)\rangle$ with the dimensions $D_n^{A}$ and $D_{N-n}^{B}$, respectively. We note that $|\varphi(i_A,i_B)\rangle$ is null whenever $n_A+n_B\neq N$, since the bases are allocated in the optimal manner described above. Therefore, once the relative position of $|\varphi(i_A,i_B)\rangle$ is settled, one can easily construct $\rho^{(n)}$ so that $\rho_n$ and $p_n$ follow successively.

Since $S_{\textrm{vN}}^{n_A=n}=S_{\textrm{vN}}^{n_B=N-n}$, in the actual calculation, the smaller side of $n_A$ and $n_B$ can be chosen such that the dimension of the obtained matrix $\rho^{(n_A)}$ or $\rho^{(n_B)}$ is smaller and easier to be diagonalized numerically. In particular, for the two channels with $n=0$ and $n=N$, there is always one side with a dimension $1$, so $S_{\textrm{vN}}^{n_A=0}=S_{\textrm{vN}}^{n_A=N}=0$ always holds. Moreover, since the relative position and the dimension of each subspace of the parts $A$ and $B$ are independent of disorder, one only needs to compute $|\varphi(i_A,i_B)\rangle$ for each sample with a given disorder strength. It is clear that the above treatment greatly enhances the computational performance by saving both the CPU hours and the memories, especially for large chain lengths.

\section{\label{sec:level1}Distributions of Local Compressibility and Density of States}
\label{sec:append_Comp}

For the clean BH model, the compressibility is relevant for describing the transition between the superfluid and the Mott insulating phases. In the presence of disorder, this quantity can be appropriately replaced by the local compressibility $K_i$ at a given energy density $\varepsilon$ as defined by \cite{40}
\begin{eqnarray}
K_i=\left\{\langle\hat n_{i}^{2}\rangle-\langle\hat n_i\rangle^2\right\}_{\textrm{av}},
\label{LCKi}
\end{eqnarray}
which essentially displays how the fluctuation of the particle number at each site is impacted by the disorder for a given $\varepsilon$. In Eq.~(\ref{LCKi}), $\left\{...\right\}_{\textrm{av}}$ stands for the average over all available disorder samples for $\varepsilon$ and $\mu$. For a given $\mu$, the eigenvalue closest to $\varepsilon$ may depend on the different samples. To enhance the statistical efficiency for the distribution of $K_i$, denoted as $P_c(K_i)$, instead of the closest eigenstate, we calculate $K_i$ for the $8$ eigenstates adjacent in eigenenergy to $\varepsilon$ for each sample. An additional average is then made over these $8$ eigenstates with a proper normalization for $P_c(K_i)$. Consequently, in view of the fact that $K_i$ measures the variance of $\hat n_i$ caused by the randomness $\mu$ at a given $\varepsilon$, one anticipates that $P_c(K_i)$ might serve as a statistics device capable of capturing the fundamentally different characteristics in the distribution of the set $\{K_i\}$ between the ETH and the MBL phases.

\begin{figure}[tb]
	\begin{center}
		\includegraphics[scale=0.6]{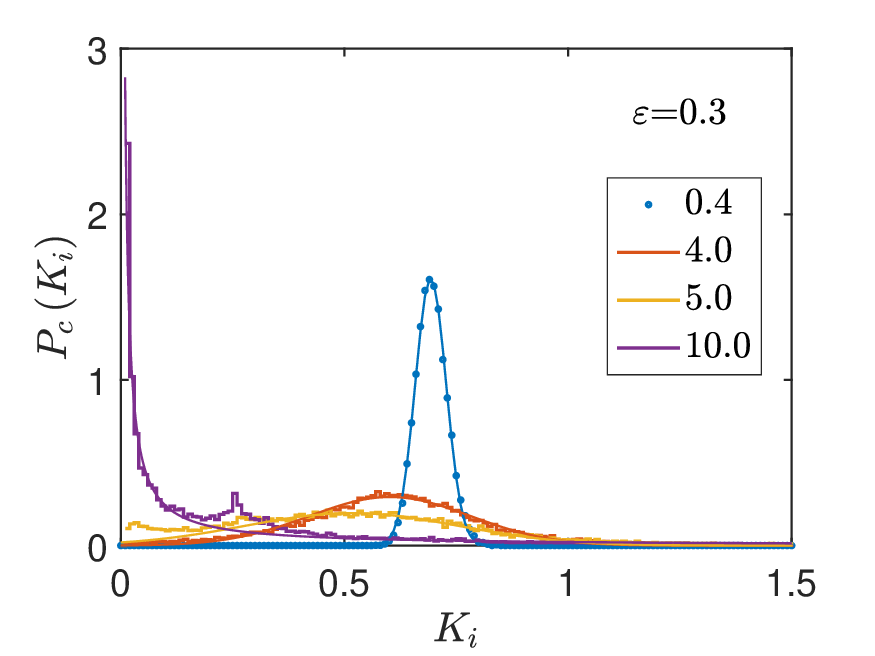}
		\caption{The distributions of the local compressibility $K_i$ at $\varepsilon=0.3$ for several disorder strengths with $L=14$ and $K_i\in[0,1.5]$. The histograms are obtained using ED and are accompanied by the fitting solid curves. For $\mu= 0.4, 4.0$, the distributions are well fitted by a Gaussian function, and a two-peak structure arises at $\mu=5.0$, where the main peak can still be fitted by a Gaussian function. When $\mu=10.0$, however, it fulfills a power law within the range of small $K_i$, meanwhile exhibiting a satellite peak at $K_i\approx0.25$.}
		\label{compre}		
	\end{center}
\end{figure}

\begin{figure}[htb]	
	\begin{center}
		\includegraphics[scale=0.6]{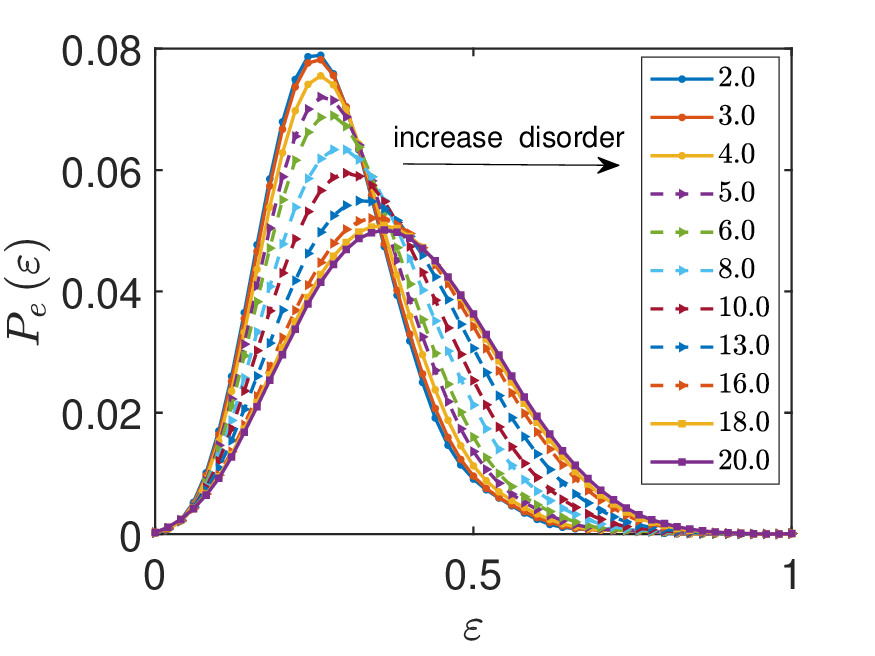}
		\caption{Energy density distributions for various disorder strengths $\mu\in[2,20]$, obtained from ED using a chain of length $L=12$.}
		\label{EnergyDensity}		
	\end{center}
\end{figure}

Figure~\ref{compre} shows $P_c(K_i)$ for $\mu=0.4,4,5,10$ at $\varepsilon=0.3$. The distribution changes from a Gaussian distribution for $\mu=0.4$ and $4$ to a power-law distribution in the large $\mu$ limit. The Gaussian distribution corresponds to the well-defined superfluid states. In general, when $\mu$ increases, the height of the curve decreases and its width gets broadened. However, as shown by $\mu=5$, while the peak is greatly broadened, a warp emerges at the small $K_i$. The warp grows rapidly with further increasing $\mu$. Eventually, it behaves as a power law shown for $\mu=10$, where there is still a small satellite peak at $K_i\approx0.25$, which would disappear when $\mu$ is sufficiently larger. Remarkably, one may find that a relatively sudden change of $P_c(K_i)$ happens between $\mu=4$ and $5$, and the disorder greatly suppresses $K_i$ in MBL. 

From the above, it's clear that in the ETH phase, the distribution is Gaussian, while in the MBL phase, it exhibits a power-law decay with a satellite peak. Notably, at $\mu=5$, the right side conforms to a Gaussian distribution, one characteristic of ETH, while the left side shows an upturn near zero and a satellite peak around 0.25, indicative of the onset of MBL. This implies that the system experiences a breakdown in thermalization at this point, with a partial MBL component, but has not yet undergone a phase transition. The phase transition appears to occur around $\mu=7.8$, as shown in Fig.~\ref{gapratio}. This implies the presence of an intermediate region in the system, being prior to the MBL phase transition. In both Figs.~\ref{gapratio} and \ref{TwoBodyCorrelation}(b), an intermediate region is observed between the yellow thermalization region's boundary and the phase transition boundary.

Research indicates that the distribution of thermalized states is positively correlated with the density of states (also called energy density distribution) \cite{schierenberg2012wigner}, with a higher density of states corresponding to a greater presence of the thermalization states. For typical spin systems, the density of states is usually approximated by a Gaussian function \cite{33}.

\begin{figure}[b]	
	\begin{center}
		\includegraphics[scale=0.6]{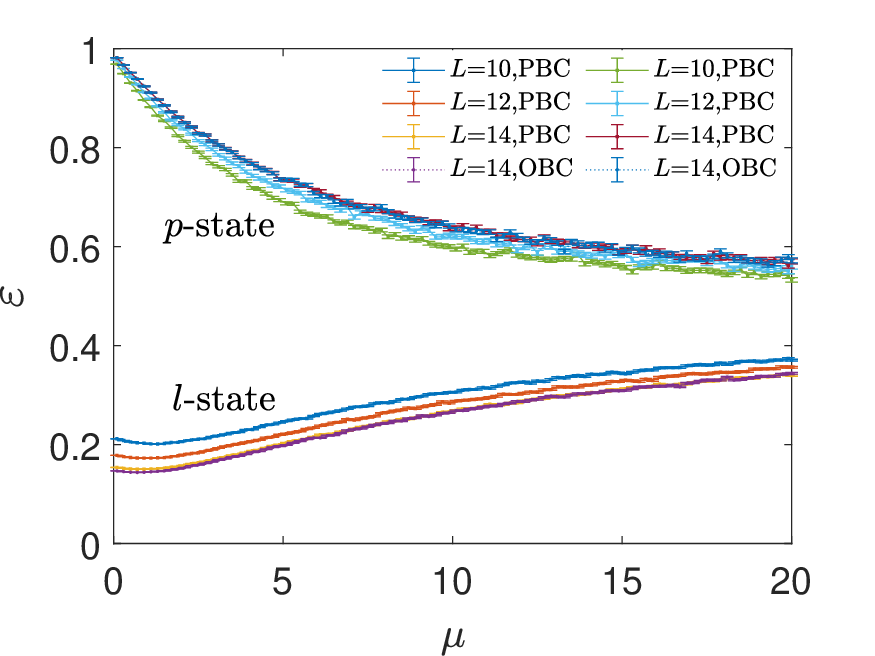}
		\caption{The average energy densities for the two kinds of initial states as a function of $\mu$ for different chain lengths. In the present study, the $p$-state corresponds to an initial state where all $L/2$ bosons are loaded at the first site and zero occupancy for the remainder of the system. At the same time, one particle occupies each site of the left part of the system, and no particle is in all sites of the right part, comprising the $l$-state.}
		\label{inienergy}		
	\end{center}	
\end{figure}

Figure~\ref{EnergyDensity} illustrates the disorder effects on the energy density distribution $P_e(\varepsilon)$ for a wide range of the disorder strengths. It is remarkable that for the Hamiltonian (\ref{dbhmodel}), its eigenstates are mainly concentrated on the range of energy densities $\varepsilon< 0.5$ for the weak disorder, where only a tiny tail shows up for $\varepsilon>0.5$. However, the distribution becomes broadened with the increasing disorder strength $\mu$. Its peak position starts to slowly shift toward $\varepsilon=0.5$ when $\mu$ is sufficiently large, where a reasonably large tail is also developed for $\varepsilon>0.5$. Indeed, under weak disorder, the thermalization states concentrate in the low-energy region, resulting in the trend of energy density distribution mainly centered around $\varepsilon < 0.5$. Unlike the spin model, we find that these $P_e(\varepsilon)$ curves can be fitted by the summation of two Gaussian functions.

\section{\label{sec:level1}Averaged Energy Density of Initial State}
\label{sec:append_Aver}

\begin{figure}[b]
	\begin{center}
		\includegraphics[scale=0.6]{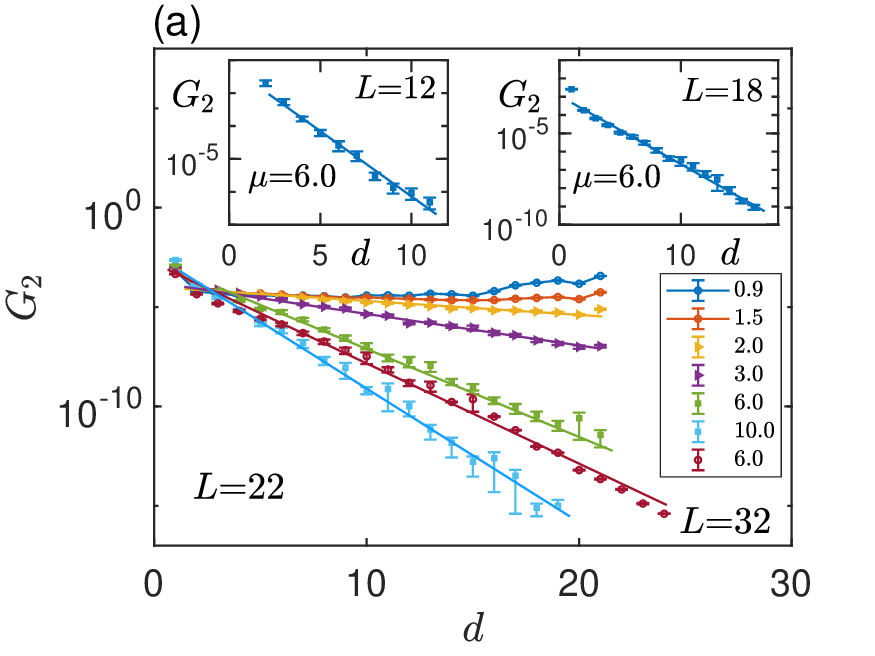}
		\includegraphics[scale=0.6]{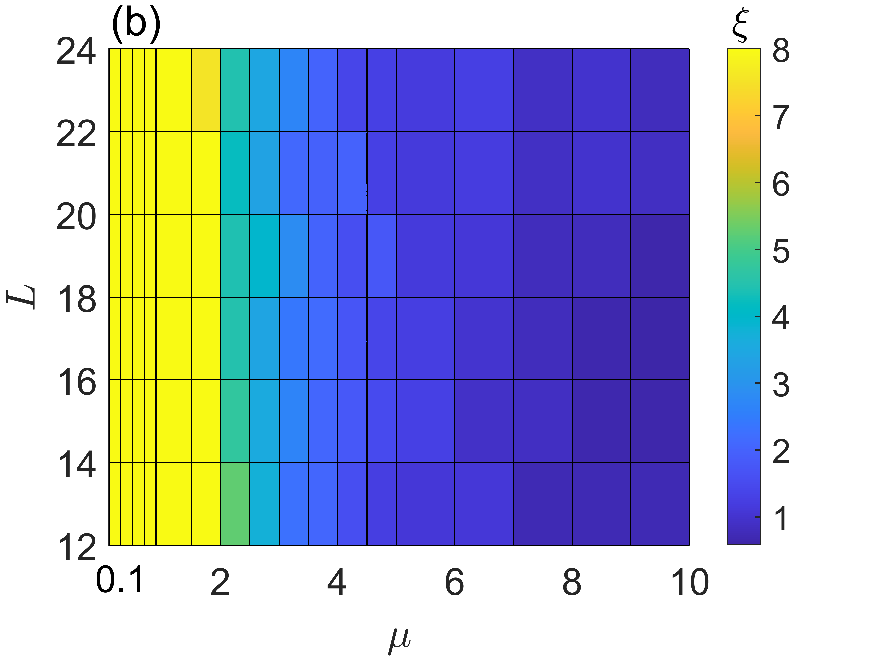}
		\caption{(a) The correlation function $G_2(d)$ for $L=22$ with $\mu=0.9,1.5,2.0,3.0,6.0,10.0$. The result for $L=32$ and $\mu=6.0$ is also included for comparison; the insets of (a) are for $G_2$ at $\mu=6.0$, left with $L=12$ and right with $L=18$. (b) The localization length $\xi$ for $L=12,14,16,18,20,22$ at different disorder strengths. The data are extracted from $G_2(d)$ evaluated in the many-body states which have been evolved up to $t=100\tau$. Here, the initial product state is chosen to be the $p$-state.}
		\label{G2_100}		
	\end{center}
\end{figure}

Figure~\ref{inienergy} shows how the averaged energy densities of low-energy $l$-state and high-energy $p$-state change as a function of the disorder strength for different chain lengths. For the $p$-state, the energy density decreases as the disorder increases. Because the system realizes cluster MBL at small disorder, the clustering tendency enhances the Hubbard interaction, leading to the high energy densities. As the disorder increases, the degree of the clustering of particles reduces, and the effect of the interaction $U$ term weakens; thus, the overall energy density decreases. Note that as the system size increases, the overall energy density also slightly grows. By contrast, it can be seen that the energy density of the $l$-state changes slowly with the increasing disorder strength and decreases with the increasing system size. Comparing to the energy density distribution shown in Fig.~\ref{EnergyDensity}, one can see that these initial states correspond to somewhat different densities of energy eigenstates. Compared with the phase diagram Fig.~\ref{gapratio}, the energy-density trajectories of these two states evolve under the change of the disorder strength, passing across several different phase regions, which explains why they have very different time-evolutionary behaviors. Using these two kinds of initial states, we present our numerical results on the respective time evolutions of the two-body density-density correlation function and the two types of entropies.

\section{\label{sec:level1}Localization Length from Steady State}
\label{sec:append_Locali}

The time evolutions of $G_2$ are calculated at $t=100\tau$ with $\tau =0.1$ \cite{35,69} for several system sizes $L$. Specifically, we perform the computations at $L=12$ using ED \cite{28}, and at $L=14,16,18$ using the global Krylov method \cite{37}, and at $L=20,22,32$ using TEBD \cite{38}, for which the $2$nd-order Trotter decomposition is implemented to suppress the systematical errors. For the TEBD calculation, $600$ samples are taken for the disorder averages. To study the single-particle localization length of cluster MBL, we take the high-energy $p$-state, i.e., all $L/2$ bosons loaded at the first site and zero occupancy for the rest of the system, which is a high-energy-density state; as the disorder strength increases, its energy density maintains in the cluster MBL region, as can be seen from Figs.~\ref{inienergy} and \ref{gapratio}. Since the first site has the highest occupancy, the occupancy number cannot be truncated in the TEBD simulations. The local physical dimension on each site is thus $L/2+1$. We note that evaluating the time evolution of $G_2$ in such a fledged Bose system without truncating the single site occupancy costs huge computational resource for moderate to large values of $L$.

Figure~\ref{G2_100}(a) shows the disorder effects on the time-evolved two-body density-density correlation function at $t=100\tau$ on a $L=22$ chain for several values of the disorder strength. The two insets therein display the results of $G_2(d)$ at $\mu=6.0$ for $L=12$ and $18$, respectively. Upon the weak disorder, i.e., in the ETH region, $G_2$ has relatively small values for the moderate values of $d$ and becomes large when $d\sim1$ or $L$. Here, the open boundary conditions are imposed. As the disorder strength gets further increased, $G_2(d)$ starts to display an exponential decrease with respect to $d$. This trend can still be visible even down to the order of $10^{-15}$, which potentially allows for the extracting of the localization length for the various cases we are interested in. In principle, using the TEBD technique, we can calculate $G_2(d)$ for relatively large sizes up to $L=32$ within our current computational capacities. This has been illustrated for the case of $\mu=6.0$ in Fig.~\ref{G2_100}(a). Note that the results up to $L=22$ have been sufficient to elucidate the time evolution of $G_2(d)$ at $t=100\tau$. We plot the localization lengths at $L=12,14,16,18,20,22$ for $\mu\in[0.1,10.0]$ in Fig.~\ref{G2_100}(b). From the color contrast of different sizes, one can see that the extracted localization lengths might be insensitive to the change in $L$ for these values of $\mu$ we consider.

For cluster MBL at small disorder, multiple particles of the Bose system can be localized at the same point. The steady-state $S_N$ and $S_C$ are thus obedient to the area law, and neither grows with time, indicating that it is a more robustly localized phenomenon \cite{84}. However, as seen from Fig.~\ref{G2_100}, the cluster MBL has a greater single-particle localization length, which seems contradictory. One possible scenario for this might be the effect of energy mismatch. Concretely, on the one hand, when part of the cluster could move, the overall probability for such processes remains low due to the creation of the significant energy mismatch. Meanwhile, on the other hand, these disentangled single particles might conversely gain the potential to hop far away in view of the fact that the disorder and the hopping generated energy offsets are barely small changes relative to the existing energy gap. Collectively, these effects give rise to a larger localization length for the single-particle process under the small disorder strengths. To some extent, the above reasoning is consistent with our explanations on the channel reflection asymmetry of $\{S_{\textrm{vN}}^{n}\}$ for the initial $p$-state as already detailed in \cite{84}.

\bibliography{ref}

\begin{thebibliography}{74}%
\makeatletter
\providecommand \@ifxundefined [1]{%
 \@ifx{#1\undefined}
}%
\providecommand \@ifnum [1]{%
 \ifnum #1\expandafter \@firstoftwo
 \else \expandafter \@secondoftwo
 \fi
}%
\providecommand \@ifx [1]{%
 \ifx #1\expandafter \@firstoftwo
 \else \expandafter \@secondoftwo
 \fi
}%
\providecommand \natexlab [1]{#1}%
\providecommand \enquote  [1]{``#1''}%
\providecommand \bibnamefont  [1]{#1}%
\providecommand \bibfnamefont [1]{#1}%
\providecommand \citenamefont [1]{#1}%
\providecommand \href@noop [0]{\@secondoftwo}%
\providecommand \href [0]{\begingroup \@sanitize@url \@href}%
\providecommand \@href[1]{\@@startlink{#1}\@@href}%
\providecommand \@@href[1]{\endgroup#1\@@endlink}%
\providecommand \@sanitize@url [0]{\catcode `\\12\catcode `\$12\catcode
  `\&12\catcode `\#12\catcode `\^12\catcode `\_12\catcode `\%12\relax}%
\providecommand \@@startlink[1]{}%
\providecommand \@@endlink[0]{}%
\providecommand \url  [0]{\begingroup\@sanitize@url \@url }%
\providecommand \@url [1]{\endgroup\@href {#1}{\urlprefix }}%
\providecommand \urlprefix  [0]{URL }%
\providecommand \Eprint [0]{\href }%
\providecommand \doibase [0]{http://dx.doi.org/}%
\providecommand \selectlanguage [0]{\@gobble}%
\providecommand \bibinfo  [0]{\@secondoftwo}%
\providecommand \bibfield  [0]{\@secondoftwo}%
\providecommand \translation [1]{[#1]}%
\providecommand \BibitemOpen [0]{}%
\providecommand \bibitemStop [0]{}%
\providecommand \bibitemNoStop [0]{.\EOS\space}%
\providecommand \EOS [0]{\spacefactor3000\relax}%
\providecommand \BibitemShut  [1]{\csname bibitem#1\endcsname}%
\let\auto@bib@innerbib\@empty
\bibitem [{\citenamefont {Anderson}(1958)}]{1}%
  \BibitemOpen
  \bibfield  {author} {\bibinfo {author} {\bibfnamefont {Philip~W}\
  \bibnamefont {Anderson}},\ }\bibfield  {title} {\enquote {\bibinfo {title}
  {Absence of diffusion in certain random lattices},}\ }\href {\doibase
  https://doi.org/10.1103/PhysRev.109.1492} {\bibfield  {journal} {\bibinfo
  {journal} {Phys. Rev.}\ }\textbf {\bibinfo {volume} {109}},\ \bibinfo {pages}
  {1492} (\bibinfo {year} {1958})}\BibitemShut {NoStop}%
\bibitem [{\citenamefont {Abrahams}\ \emph {et~al.}(1979)\citenamefont
  {Abrahams}, \citenamefont {Anderson}, \citenamefont {Licciardello} \emph
  {et~al.}}]{44}%
  \BibitemOpen
  \bibfield  {author} {\bibinfo {author} {\bibfnamefont {Elihu}\ \bibnamefont
  {Abrahams}}, \bibinfo {author} {\bibfnamefont {P.~W.}\ \bibnamefont
  {Anderson}}, \bibinfo {author} {\bibfnamefont {D.~C.}\ \bibnamefont
  {Licciardello}},  \emph {et~al.},\ }\bibfield  {title} {\enquote {\bibinfo
  {title} {Scaling theory of localization: Absence of quantum diffusion in two
  dimensions},}\ }\href {\doibase https://doi.org/10.1103/PhysRevLett.42.673}
  {\bibfield  {journal} {\bibinfo  {journal} {Phys. Rev. Lett.}\ }\textbf
  {\bibinfo {volume} {42}},\ \bibinfo {pages} {673} (\bibinfo {year}
  {1979})}\BibitemShut {NoStop}%
\bibitem [{\citenamefont {Vollhardt}\ and\ \citenamefont {Wölfle}(1980)}]{45}%
  \BibitemOpen
  \bibfield  {author} {\bibinfo {author} {\bibfnamefont {D.}~\bibnamefont
  {Vollhardt}}\ and\ \bibinfo {author} {\bibfnamefont {P.}~\bibnamefont
  {Wölfle}},\ }\bibfield  {title} {\enquote {\bibinfo {title} {Diagrammatic,
  self-consistent treatment of the anderson localization problem in $d\le 2$
  dimensions},}\ }\href {\doibase https://doi.org/10.1103/PhysRevB.22.4666}
  {\bibfield  {journal} {\bibinfo  {journal} {Phys. Rev. B}\ }\textbf {\bibinfo
  {volume} {22}},\ \bibinfo {pages} {4666} (\bibinfo {year}
  {1980})}\BibitemShut {NoStop}%
\bibitem [{\citenamefont {John}(1984)}]{46}%
  \BibitemOpen
  \bibfield  {author} {\bibinfo {author} {\bibfnamefont {Sajeev}\ \bibnamefont
  {John}},\ }\bibfield  {title} {\enquote {\bibinfo {title} {Electromagnetic
  absorption in a disordered medium near a photon mobility edge},}\ }\href
  {\doibase https://doi.org/10.1103/PhysRevLett.53.2169} {\bibfield  {journal}
  {\bibinfo  {journal} {Phys. Rev. Lett.}\ }\textbf {\bibinfo {volume} {53}},\
  \bibinfo {pages} {2169} (\bibinfo {year} {1984})}\BibitemShut {NoStop}%
\bibitem [{\citenamefont {Arya}\ \emph {et~al.}(1985)\citenamefont {Arya},
  \citenamefont {Su},\ and\ \citenamefont {Birman}}]{47}%
  \BibitemOpen
  \bibfield  {author} {\bibinfo {author} {\bibfnamefont {K.}~\bibnamefont
  {Arya}}, \bibinfo {author} {\bibfnamefont {Z.~B.}\ \bibnamefont {Su}}, \ and\
  \bibinfo {author} {\bibfnamefont {Joseph~L.}\ \bibnamefont {Birman}},\
  }\bibfield  {title} {\enquote {\bibinfo {title} {Localization of the surface
  plasmon polariton caused by random roughness and its role in surface-enhanced
  optical phenomena},}\ }\href {\doibase
  https://doi.org/10.1103/PhysRevLett.54.1559} {\bibfield  {journal} {\bibinfo
  {journal} {Phys. Rev. Lett.}\ }\textbf {\bibinfo {volume} {54}},\ \bibinfo
  {pages} {1559} (\bibinfo {year} {1985})}\BibitemShut {NoStop}%
\bibitem [{\citenamefont {Chu}\ and\ \citenamefont {Zhang}(1988)}]{48}%
  \BibitemOpen
  \bibfield  {author} {\bibinfo {author} {\bibfnamefont {Qian-Jin}\
  \bibnamefont {Chu}}\ and\ \bibinfo {author} {\bibfnamefont {Zhao-Qing}\
  \bibnamefont {Zhang}},\ }\bibfield  {title} {\enquote {\bibinfo {title}
  {Localization of phonons in mixed crystals},}\ }\href {\doibase
  https://doi.org/10.1103/PhysRevB.38.4906} {\bibfield  {journal} {\bibinfo
  {journal} {Phys. Rev. B}\ }\textbf {\bibinfo {volume} {38}},\ \bibinfo
  {pages} {4906} (\bibinfo {year} {1988})}\BibitemShut {NoStop}%
\bibitem [{\citenamefont {Nandkishore}\ and\ \citenamefont {Huse}(2015)}]{2}%
  \BibitemOpen
  \bibfield  {author} {\bibinfo {author} {\bibfnamefont {Rahul}\ \bibnamefont
  {Nandkishore}}\ and\ \bibinfo {author} {\bibfnamefont {David~A.}\
  \bibnamefont {Huse}},\ }\bibfield  {title} {\enquote {\bibinfo {title}
  {Many-body localization and thermalization in quantum statistical
  mechanics},}\ }\href
  {https://www.annualreviews.org/doi/abs/10.1146/annurev-conmatphys-031214-014726}
  {\bibfield  {journal} {\bibinfo  {journal} {Annu. Rev. Condens. Matter
  Phys.}\ }\textbf {\bibinfo {volume} {6}},\ \bibinfo {pages} {15–38}
  (\bibinfo {year} {2015})}\BibitemShut {NoStop}%
\bibitem [{\citenamefont {Basko}\ \emph {et~al.}(2006)\citenamefont {Basko},
  \citenamefont {Aleiner},\ and\ \citenamefont {Altshuler}}]{53}%
  \BibitemOpen
  \bibfield  {author} {\bibinfo {author} {\bibfnamefont {Denis~M.}\
  \bibnamefont {Basko}}, \bibinfo {author} {\bibfnamefont {Igor~L.}\
  \bibnamefont {Aleiner}}, \ and\ \bibinfo {author} {\bibfnamefont {Boris~L.}\
  \bibnamefont {Altshuler}},\ }\bibfield  {title} {\enquote {\bibinfo {title}
  {Metal--insulator transition in a weakly interacting many-electron system
  with localized single-particle states},}\ }\href {\doibase
  https://doi.org/10.1016/j.aop.2005.11.014} {\bibfield  {journal} {\bibinfo
  {journal} {Ann. Phys.}\ }\textbf {\bibinfo {volume} {321}},\ \bibinfo {pages}
  {1126} (\bibinfo {year} {2006})}\BibitemShut {NoStop}%
\bibitem [{\citenamefont {Oganesyan}\ and\ \citenamefont {Huse}(2007)}]{54}%
  \BibitemOpen
  \bibfield  {author} {\bibinfo {author} {\bibfnamefont {V.}~\bibnamefont
  {Oganesyan}}\ and\ \bibinfo {author} {\bibfnamefont {D.~A.}\ \bibnamefont
  {Huse}},\ }\bibfield  {title} {\enquote {\bibinfo {title} {Localization of
  interacting fermions at high temperature},}\ }\href {\doibase
  https://doi.org/10.1103/PhysRevB.75.155111} {\bibfield  {journal} {\bibinfo
  {journal} {Phys. Rev. B}\ }\textbf {\bibinfo {volume} {75}},\ \bibinfo
  {pages} {155111} (\bibinfo {year} {2007})}\BibitemShut {NoStop}%
\bibitem [{\citenamefont {Žnidarič}\ \emph {et~al.}(2008)\citenamefont
  {Žnidarič}, \citenamefont {Prosen},\ and\ \citenamefont {Prelovšek}}]{55}%
  \BibitemOpen
  \bibfield  {author} {\bibinfo {author} {\bibfnamefont {Marko}\ \bibnamefont
  {Žnidarič}}, \bibinfo {author} {\bibfnamefont {Tomaž}\ \bibnamefont
  {Prosen}}, \ and\ \bibinfo {author} {\bibfnamefont {Peter}\ \bibnamefont
  {Prelovšek}},\ }\bibfield  {title} {\enquote {\bibinfo {title} {Many-body
  localization in the heisenberg xxz magnet in a random field},}\ }\href
  {\doibase https://doi.org/10.1103/PhysRevB.77.064426} {\bibfield  {journal}
  {\bibinfo  {journal} {Phys. Rev. B}\ }\textbf {\bibinfo {volume} {77}},\
  \bibinfo {pages} {064426} (\bibinfo {year} {2008})}\BibitemShut {NoStop}%
\bibitem [{\citenamefont {Deutsch}(1991)}]{56}%
  \BibitemOpen
  \bibfield  {author} {\bibinfo {author} {\bibfnamefont {J.~M.}\ \bibnamefont
  {Deutsch}},\ }\bibfield  {title} {\enquote {\bibinfo {title} {Quantum
  statistical mechanics in a closed system},}\ }\href {\doibase
  https://doi.org/10.1103/PhysRevA.43.2046} {\bibfield  {journal} {\bibinfo
  {journal} {Phys. Rev. A}\ }\textbf {\bibinfo {volume} {43}},\ \bibinfo
  {pages} {2046} (\bibinfo {year} {1991})}\BibitemShut {NoStop}%
\bibitem [{\citenamefont {Srednicki}(1994)}]{57}%
  \BibitemOpen
  \bibfield  {author} {\bibinfo {author} {\bibfnamefont {Mark}\ \bibnamefont
  {Srednicki}},\ }\bibfield  {title} {\enquote {\bibinfo {title} {Chaos and
  quantum thermalization},}\ }\href {\doibase
  https://doi.org/10.1103/PhysRevE.50.888} {\bibfield  {journal} {\bibinfo
  {journal} {Phys. Rev. E}\ }\textbf {\bibinfo {volume} {50}},\ \bibinfo
  {pages} {888} (\bibinfo {year} {1994})}\BibitemShut {NoStop}%
\bibitem [{\citenamefont {Serbyn}\ \emph
  {et~al.}(2013{\natexlab{a}})\citenamefont {Serbyn}, \citenamefont {Papić},\
  and\ \citenamefont {Abanin}}]{4}%
  \BibitemOpen
  \bibfield  {author} {\bibinfo {author} {\bibfnamefont {Maksym}\ \bibnamefont
  {Serbyn}}, \bibinfo {author} {\bibfnamefont {Z.}~\bibnamefont {Papić}}, \
  and\ \bibinfo {author} {\bibfnamefont {Dmitry~A.}\ \bibnamefont {Abanin}},\
  }\bibfield  {title} {\enquote {\bibinfo {title} {Local conservation laws and
  the structure of the many-body localized states},}\ }\href {\doibase
  https://doi.org/10.1103/PhysRevLett.111.127201} {\bibfield  {journal}
  {\bibinfo  {journal} {Phys. Rev. Lett.}\ }\textbf {\bibinfo {volume} {111}},\
  \bibinfo {pages} {127201} (\bibinfo {year} {2013}{\natexlab{a}})}\BibitemShut
  {NoStop}%
\bibitem [{\citenamefont {Chandran}\ \emph {et~al.}(2015)\citenamefont
  {Chandran}, \citenamefont {Kim}, \citenamefont {Vidal} \emph {et~al.}}]{5}%
  \BibitemOpen
  \bibfield  {author} {\bibinfo {author} {\bibfnamefont {Anushya}\ \bibnamefont
  {Chandran}}, \bibinfo {author} {\bibfnamefont {Isaac~H.}\ \bibnamefont
  {Kim}}, \bibinfo {author} {\bibfnamefont {Guifre}\ \bibnamefont {Vidal}},
  \emph {et~al.},\ }\bibfield  {title} {\enquote {\bibinfo {title}
  {Constructing local integrals of motion in the many-body localized phase},}\
  }\href {\doibase https://doi.org/10.1103/PhysRevB.91.085425} {\bibfield
  {journal} {\bibinfo  {journal} {Phys. Rev. B}\ }\textbf {\bibinfo {volume}
  {91}},\ \bibinfo {pages} {085425} (\bibinfo {year} {2015})}\BibitemShut
  {NoStop}%
\bibitem [{\citenamefont {Geraedts}\ \emph {et~al.}(2017)\citenamefont
  {Geraedts}, \citenamefont {Bhatt},\ and\ \citenamefont {Nandkishore}}]{6}%
  \BibitemOpen
  \bibfield  {author} {\bibinfo {author} {\bibfnamefont {Scott~D.}\
  \bibnamefont {Geraedts}}, \bibinfo {author} {\bibfnamefont {R.~N.}\
  \bibnamefont {Bhatt}}, \ and\ \bibinfo {author} {\bibfnamefont {Rahul}\
  \bibnamefont {Nandkishore}},\ }\bibfield  {title} {\enquote {\bibinfo {title}
  {Emergent local integrals of motion without a complete set of localized
  eigenstates},}\ }\href {\doibase https://doi.org/10.1103/PhysRevB.95.064204}
  {\bibfield  {journal} {\bibinfo  {journal} {Phys. Rev. B}\ }\textbf {\bibinfo
  {volume} {95}},\ \bibinfo {pages} {064204} (\bibinfo {year}
  {2017})}\BibitemShut {NoStop}%
\bibitem [{\citenamefont {D’Alessio}\ \emph {et~al.}(2016)\citenamefont
  {D’Alessio}, \citenamefont {Kafri}, \citenamefont {Polkovnikov} \emph
  {et~al.}}]{3}%
  \BibitemOpen
  \bibfield  {author} {\bibinfo {author} {\bibfnamefont {Luca}\ \bibnamefont
  {D’Alessio}}, \bibinfo {author} {\bibfnamefont {Yariv}\ \bibnamefont
  {Kafri}}, \bibinfo {author} {\bibfnamefont {Anatoli}\ \bibnamefont
  {Polkovnikov}},  \emph {et~al.},\ }\bibfield  {title} {\enquote {\bibinfo
  {title} {From quantum chaos and eigenstate thermalization to statistical
  mechanics and thermodynamics},}\ }\href {\doibase
  https://doi.org/10.1080/00018732.2016.1198134} {\bibfield  {journal}
  {\bibinfo  {journal} {Adv. in Phys.}\ }\textbf {\bibinfo {volume} {65}},\
  \bibinfo {pages} {3,239–362} (\bibinfo {year} {2016})}\BibitemShut
  {NoStop}%
\bibitem [{\citenamefont {Lazarides}\ \emph {et~al.}(2015)\citenamefont
  {Lazarides}, \citenamefont {Das},\ and\ \citenamefont {Moessner}}]{14}%
  \BibitemOpen
  \bibfield  {author} {\bibinfo {author} {\bibfnamefont {Achilleas}\
  \bibnamefont {Lazarides}}, \bibinfo {author} {\bibfnamefont {Arnab}\
  \bibnamefont {Das}}, \ and\ \bibinfo {author} {\bibfnamefont {Roderich}\
  \bibnamefont {Moessner}},\ }\href {\doibase
  https://doi.org/10.1103/PhysRevLett.115.030402} {\bibfield  {journal}
  {\bibinfo  {journal} {Phys. Rev. Lett.}\ }\textbf {\bibinfo {volume} {115}},\
  \bibinfo {pages} {030402} (\bibinfo {year} {2015})}\BibitemShut {NoStop}%
\bibitem [{\citenamefont {Ponte}\ \emph {et~al.}(2015)\citenamefont {Ponte},
  \citenamefont {Papić}, \citenamefont {Huveneers} \emph {et~al.}}]{15}%
  \BibitemOpen
  \bibfield  {author} {\bibinfo {author} {\bibfnamefont {Pedro}\ \bibnamefont
  {Ponte}}, \bibinfo {author} {\bibfnamefont {Z.}~\bibnamefont {Papić}},
  \bibinfo {author} {\bibfnamefont {François}\ \bibnamefont {Huveneers}},
  \emph {et~al.},\ }\bibfield  {title} {\enquote {\bibinfo {title} {Fate of
  many-body localization under periodic driving},}\ }\href {\doibase
  https://doi.org/10.1103/PhysRevLett.114.140401} {\bibfield  {journal}
  {\bibinfo  {journal} {Phys. Rev. Lett.}\ }\textbf {\bibinfo {volume} {114}},\
  \bibinfo {pages} {140401} (\bibinfo {year} {2015})}\BibitemShut {NoStop}%
\bibitem [{\citenamefont {Else}\ \emph {et~al.}(2016)\citenamefont {Else},
  \citenamefont {Bauer},\ and\ \citenamefont {Nayak}}]{16}%
  \BibitemOpen
  \bibfield  {author} {\bibinfo {author} {\bibfnamefont {Dominic~V.}\
  \bibnamefont {Else}}, \bibinfo {author} {\bibfnamefont {Bela}\ \bibnamefont
  {Bauer}}, \ and\ \bibinfo {author} {\bibfnamefont {Chetan}\ \bibnamefont
  {Nayak}},\ }\bibfield  {title} {\enquote {\bibinfo {title} {Floquet time
  crystals},}\ }\href {\doibase https://doi.org/10.1103/PhysRevLett.117.090402}
  {\bibfield  {journal} {\bibinfo  {journal} {Phys. Rev. Lett.}\ }\textbf
  {\bibinfo {volume} {117}},\ \bibinfo {pages} {090402} (\bibinfo {year}
  {2016})}\BibitemShut {NoStop}%
\bibitem [{\citenamefont {Yao}\ \emph {et~al.}(2017)\citenamefont {Yao},
  \citenamefont {Potter}, \citenamefont {Potirniche} \emph {et~al.}}]{17}%
  \BibitemOpen
  \bibfield  {author} {\bibinfo {author} {\bibfnamefont {N.~Y.}\ \bibnamefont
  {Yao}}, \bibinfo {author} {\bibfnamefont {A.~C.}\ \bibnamefont {Potter}},
  \bibinfo {author} {\bibfnamefont {I.-D.}\ \bibnamefont {Potirniche}},  \emph
  {et~al.},\ }\bibfield  {title} {\enquote {\bibinfo {title} {Discrete time
  crystals: Rigidity, criticality, and realizations},}\ }\href {\doibase
  https://doi.org/10.1103/PhysRevLett.118.030401} {\bibfield  {journal}
  {\bibinfo  {journal} {Phys. Rev. Lett.}\ }\textbf {\bibinfo {volume} {118}},\
  \bibinfo {pages} {030401} (\bibinfo {year} {2017})}\BibitemShut {NoStop}%
\bibitem [{\citenamefont {Zhang}\ \emph {et~al.}(2017)\citenamefont {Zhang},
  \citenamefont {Hess}, \citenamefont {Kyprianidis} \emph {et~al.}}]{18}%
  \BibitemOpen
  \bibfield  {author} {\bibinfo {author} {\bibfnamefont {J.}~\bibnamefont
  {Zhang}}, \bibinfo {author} {\bibfnamefont {P.~W.}\ \bibnamefont {Hess}},
  \bibinfo {author} {\bibfnamefont {A.}~\bibnamefont {Kyprianidis}},  \emph
  {et~al.},\ }\bibfield  {title} {\enquote {\bibinfo {title} {Observation of a
  discrete time crystal},}\ }\href
  {https://www.nature.com/articles/nature21413} {\bibfield  {journal} {\bibinfo
   {journal} {Nature}\ }\textbf {\bibinfo {volume} {543}},\ \bibinfo {pages}
  {217–220} (\bibinfo {year} {2017})}\BibitemShut {NoStop}%
\bibitem [{\citenamefont {Choi}\ \emph {et~al.}(2017)\citenamefont {Choi},
  \citenamefont {Choi}, \citenamefont {Landig} \emph {et~al.}}]{19}%
  \BibitemOpen
  \bibfield  {author} {\bibinfo {author} {\bibfnamefont {Soonwon}\ \bibnamefont
  {Choi}}, \bibinfo {author} {\bibfnamefont {Joonhee}\ \bibnamefont {Choi}},
  \bibinfo {author} {\bibfnamefont {Renate}\ \bibnamefont {Landig}},  \emph
  {et~al.},\ }\bibfield  {title} {\enquote {\bibinfo {title} {Observation of
  discrete time-crystalline order in a disordered dipolar many-body system},}\
  }\href {\doibase https://doi.org/10.1038/nature21426} {\bibfield  {journal}
  {\bibinfo  {journal} {Nature}\ }\textbf {\bibinfo {volume} {543}},\ \bibinfo
  {pages} {221–225} (\bibinfo {year} {2017})}\BibitemShut {NoStop}%
\bibitem [{\citenamefont {Kjäll}\ \emph {et~al.}(2014)\citenamefont {Kjäll},
  \citenamefont {Bardarson},\ and\ \citenamefont {Pollmann}}]{13}%
  \BibitemOpen
  \bibfield  {author} {\bibinfo {author} {\bibfnamefont {Jonas~A.}\
  \bibnamefont {Kjäll}}, \bibinfo {author} {\bibfnamefont {Jens~H.}\
  \bibnamefont {Bardarson}}, \ and\ \bibinfo {author} {\bibfnamefont {Frank}\
  \bibnamefont {Pollmann}},\ }\bibfield  {title} {\enquote {\bibinfo {title}
  {Many-body localization in a disordered quantum ising chain},}\ }\href
  {\doibase https://doi.org/10.1103/PhysRevLett.113.107204} {\bibfield
  {journal} {\bibinfo  {journal} {Phys. Rev. Lett.}\ }\textbf {\bibinfo
  {volume} {113}},\ \bibinfo {pages} {107204} (\bibinfo {year}
  {2014})}\BibitemShut {NoStop}%
\bibitem [{\citenamefont {Brenes}\ \emph {et~al.}(2018)\citenamefont {Brenes},
  \citenamefont {Dalmonte}, \citenamefont {Heyl} \emph {et~al.}}]{22}%
  \BibitemOpen
  \bibfield  {author} {\bibinfo {author} {\bibfnamefont {Marlon}\ \bibnamefont
  {Brenes}}, \bibinfo {author} {\bibfnamefont {Marcello}\ \bibnamefont
  {Dalmonte}}, \bibinfo {author} {\bibfnamefont {Markus}\ \bibnamefont {Heyl}},
   \emph {et~al.},\ }\bibfield  {title} {\enquote {\bibinfo {title} {Many-body
  localization dynamics from gauge invariance},}\ }\href {\doibase
  https://doi.org/10.1103/PhysRevLett.120.030601} {\bibfield  {journal}
  {\bibinfo  {journal} {Phys. Rev. Lett.}\ }\textbf {\bibinfo {volume} {120}},\
  \bibinfo {pages} {030601} (\bibinfo {year} {2018})}\BibitemShut {NoStop}%
\bibitem [{\citenamefont {Levi}\ \emph {et~al.}(2016)\citenamefont {Levi},
  \citenamefont {Heyl}, \citenamefont {Lesanovsky} \emph {et~al.}}]{58}%
  \BibitemOpen
  \bibfield  {author} {\bibinfo {author} {\bibfnamefont {Emanuele}\
  \bibnamefont {Levi}}, \bibinfo {author} {\bibfnamefont {Markus}\ \bibnamefont
  {Heyl}}, \bibinfo {author} {\bibfnamefont {Igor}\ \bibnamefont {Lesanovsky}},
   \emph {et~al.},\ }\bibfield  {title} {\enquote {\bibinfo {title} {Robustness
  of many-body localization in the presence of dissipation},}\ }\href {\doibase
  https://doi.org/10.1103/PhysRevLett.116.237203} {\bibfield  {journal}
  {\bibinfo  {journal} {Phys. Rev. Lett.}\ }\textbf {\bibinfo {volume} {116}},\
  \bibinfo {pages} {237203} (\bibinfo {year} {2016})}\BibitemShut {NoStop}%
\bibitem [{\citenamefont {Fischer}\ \emph {et~al.}(2016)\citenamefont
  {Fischer}, \citenamefont {Maksymenko},\ and\ \citenamefont {Altman}}]{59}%
  \BibitemOpen
  \bibfield  {author} {\bibinfo {author} {\bibfnamefont {Mark~H.}\ \bibnamefont
  {Fischer}}, \bibinfo {author} {\bibfnamefont {Mykola}\ \bibnamefont
  {Maksymenko}}, \ and\ \bibinfo {author} {\bibfnamefont {Ehud}\ \bibnamefont
  {Altman}},\ }\bibfield  {title} {\enquote {\bibinfo {title} {Dynamics of a
  many-body-localized system coupled to a bath},}\ }\href {\doibase
  https://doi.org/10.1103/PhysRevLett.116.160401} {\bibfield  {journal}
  {\bibinfo  {journal} {Phys. Rev. Lett.}\ }\textbf {\bibinfo {volume} {116}},\
  \bibinfo {pages} {160401} (\bibinfo {year} {2016})}\BibitemShut {NoStop}%
\bibitem [{\citenamefont {Lüschen}\ \emph {et~al.}(2017)\citenamefont
  {Lüschen}, \citenamefont {Bordia}, \citenamefont {Hodgman} \emph
  {et~al.}}]{60}%
  \BibitemOpen
  \bibfield  {author} {\bibinfo {author} {\bibfnamefont {Henrik~P.}\
  \bibnamefont {Lüschen}}, \bibinfo {author} {\bibfnamefont {Pranjal}\
  \bibnamefont {Bordia}}, \bibinfo {author} {\bibfnamefont {Sean~S.}\
  \bibnamefont {Hodgman}},  \emph {et~al.},\ }\bibfield  {title} {\enquote
  {\bibinfo {title} {Signatures of many-body localization in a controlled open
  quantum system},}\ }\href {\doibase
  https://doi.org/10.1103/PhysRevX.7.011034} {\bibfield  {journal} {\bibinfo
  {journal} {Phys. Rev. X}\ }\textbf {\bibinfo {volume} {7}},\ \bibinfo {pages}
  {011034} (\bibinfo {year} {2017})}\BibitemShut {NoStop}%
\bibitem [{\citenamefont {Ren}\ \emph {et~al.}(2020)\citenamefont {Ren},
  \citenamefont {Li}, \citenamefont {Li}, \citenamefont {Cai},\ and\
  \citenamefont {Wang}}]{61}%
  \BibitemOpen
  \bibfield  {author} {\bibinfo {author} {\bibfnamefont {Jie}\ \bibnamefont
  {Ren}}, \bibinfo {author} {\bibfnamefont {Qiaoyi}\ \bibnamefont {Li}},
  \bibinfo {author} {\bibfnamefont {Wei}\ \bibnamefont {Li}}, \bibinfo {author}
  {\bibfnamefont {Zi}~\bibnamefont {Cai}}, \ and\ \bibinfo {author}
  {\bibfnamefont {Xiaoqun}\ \bibnamefont {Wang}},\ }\bibfield  {title}
  {\enquote {\bibinfo {title} {Noise-driven universal dynamics towards an
  infinite temperature state},}\ }\href {\doibase
  https://doi.org/10.1103/PhysRevLett.124.130602} {\bibfield  {journal}
  {\bibinfo  {journal} {Phys. Rev. Lett.}\ }\textbf {\bibinfo {volume} {124}},\
  \bibinfo {pages} {130602} (\bibinfo {year} {2020})}\BibitemShut {NoStop}%
\bibitem [{\citenamefont {Žnidarič}\ \emph {et~al.}(2016)\citenamefont
  {Žnidarič}, \citenamefont {Scardicchio},\ and\ \citenamefont {Varma}}]{62}%
  \BibitemOpen
  \bibfield  {author} {\bibinfo {author} {\bibfnamefont {Marko}\ \bibnamefont
  {Žnidarič}}, \bibinfo {author} {\bibfnamefont {Antonello}\ \bibnamefont
  {Scardicchio}}, \ and\ \bibinfo {author} {\bibfnamefont {Vipin~Kerala}\
  \bibnamefont {Varma}},\ }\bibfield  {title} {\enquote {\bibinfo {title}
  {Diffusive and subdiffusive spin transport in the ergodic phase of a
  many-body localizable system},}\ }\href {\doibase
  https://doi.org/10.1103/PhysRevLett.117.040601} {\bibfield  {journal}
  {\bibinfo  {journal} {Phys. Rev. Lett.}\ }\textbf {\bibinfo {volume} {117}},\
  \bibinfo {pages} {040601} (\bibinfo {year} {2016})}\BibitemShut {NoStop}%
\bibitem [{\citenamefont {Kelly}\ \emph {et~al.}(2020)\citenamefont {Kelly},
  \citenamefont {Nandkishore},\ and\ \citenamefont {Marino}}]{75}%
  \BibitemOpen
  \bibfield  {author} {\bibinfo {author} {\bibfnamefont {Shane~P.}\
  \bibnamefont {Kelly}}, \bibinfo {author} {\bibfnamefont {Rahul}\ \bibnamefont
  {Nandkishore}}, \ and\ \bibinfo {author} {\bibfnamefont {Jamir}\ \bibnamefont
  {Marino}},\ }\bibfield  {title} {\enquote {\bibinfo {title} {Exploring
  many-body localization in quantum systems coupled to an environment via
  wegner-wilson flows},}\ }\href {\doibase
  https://doi.org/10.1016/j.nuclphysb.2019.114886} {\bibfield  {journal}
  {\bibinfo  {journal} {Nucl. Phys. B}\ }\textbf {\bibinfo {volume} {951}},\
  \bibinfo {pages} {114886} (\bibinfo {year} {2020})}\BibitemShut {NoStop}%
\bibitem [{\citenamefont {Chamon}\ \emph {et~al.}(2014)\citenamefont {Chamon},
  \citenamefont {Hamma},\ and\ \citenamefont {Mucciolo}}]{9}%
  \BibitemOpen
  \bibfield  {author} {\bibinfo {author} {\bibfnamefont {Claudio}\ \bibnamefont
  {Chamon}}, \bibinfo {author} {\bibfnamefont {Alioscia}\ \bibnamefont
  {Hamma}}, \ and\ \bibinfo {author} {\bibfnamefont {Eduardo~R.}\ \bibnamefont
  {Mucciolo}},\ }\bibfield  {title} {\enquote {\bibinfo {title} {Emergent
  irreversibility and entanglement spectrum statistics},}\ }\href {\doibase
  https://doi.org/10.1103/PhysRevLett.112.240501} {\bibfield  {journal}
  {\bibinfo  {journal} {Phys. Rev. Lett.}\ }\textbf {\bibinfo {volume} {112}},\
  \bibinfo {pages} {240501} (\bibinfo {year} {2014})}\BibitemShut {NoStop}%
\bibitem [{\citenamefont {Laumann}\ \emph {et~al.}(2014)\citenamefont
  {Laumann}, \citenamefont {Pal},\ and\ \citenamefont {Scardicchio}}]{10}%
  \BibitemOpen
  \bibfield  {author} {\bibinfo {author} {\bibfnamefont {Christopher~R}\
  \bibnamefont {Laumann}}, \bibinfo {author} {\bibfnamefont {A}~\bibnamefont
  {Pal}}, \ and\ \bibinfo {author} {\bibfnamefont {A}~\bibnamefont
  {Scardicchio}},\ }\bibfield  {title} {\enquote {\bibinfo {title} {Many-body
  mobility edge in a mean-field quantum spin glass},}\ }\href {\doibase
  https://doi.org/10.1103/PhysRevLett.113.200405} {\bibfield  {journal}
  {\bibinfo  {journal} {Phys. Rev. Lett.}\ }\textbf {\bibinfo {volume} {113}},\
  \bibinfo {pages} {200405} (\bibinfo {year} {2014})}\BibitemShut {NoStop}%
\bibitem [{\citenamefont {Luitz}\ \emph {et~al.}(2015)\citenamefont {Luitz},
  \citenamefont {Laflorencie},\ and\ \citenamefont {Alet}}]{11}%
  \BibitemOpen
  \bibfield  {author} {\bibinfo {author} {\bibfnamefont {David~J.}\
  \bibnamefont {Luitz}}, \bibinfo {author} {\bibfnamefont {Nicolas}\
  \bibnamefont {Laflorencie}}, \ and\ \bibinfo {author} {\bibfnamefont
  {Fabien}\ \bibnamefont {Alet}},\ }\bibfield  {title} {\enquote {\bibinfo
  {title} {Many-body localization edge in the random-field heisenberg chain},}\
  }\href {\doibase https://doi.org/10.1103/PhysRevB.91.081103} {\bibfield
  {journal} {\bibinfo  {journal} {Phys. Rev. B}\ }\textbf {\bibinfo {volume}
  {91}},\ \bibinfo {pages} {081103(R)} (\bibinfo {year} {2015})}\BibitemShut
  {NoStop}%
\bibitem [{\citenamefont {Mondragon-Shem}\ \emph {et~al.}(2015)\citenamefont
  {Mondragon-Shem}, \citenamefont {Pal}, \citenamefont {Hughes} \emph
  {et~al.}}]{12}%
  \BibitemOpen
  \bibfield  {author} {\bibinfo {author} {\bibfnamefont {Ian}\ \bibnamefont
  {Mondragon-Shem}}, \bibinfo {author} {\bibfnamefont {Arijeet}\ \bibnamefont
  {Pal}}, \bibinfo {author} {\bibfnamefont {Taylor~L.}\ \bibnamefont {Hughes}},
   \emph {et~al.},\ }\bibfield  {title} {\enquote {\bibinfo {title} {Many-body
  mobility edge due to symmetry-constrained dynamics and strong
  interactions},}\ }\href {\doibase https://doi.org/10.1103/PhysRevB.92.064203}
  {\bibfield  {journal} {\bibinfo  {journal} {Phys. Rev. B}\ }\textbf {\bibinfo
  {volume} {92}},\ \bibinfo {pages} {064203} (\bibinfo {year}
  {2015})}\BibitemShut {NoStop}%
\bibitem [{\citenamefont {Mondaini}\ and\ \citenamefont {Cai}(2017)}]{29}%
  \BibitemOpen
  \bibfield  {author} {\bibinfo {author} {\bibfnamefont {Rubem}\ \bibnamefont
  {Mondaini}}\ and\ \bibinfo {author} {\bibfnamefont {Zi}~\bibnamefont {Cai}},\
  }\bibfield  {title} {\enquote {\bibinfo {title} {Many-body self-localization
  in a translation-invariant hamiltonian},}\ }\href {\doibase
  https://doi.org/10.1103/PhysRevB.96.035153} {\bibfield  {journal} {\bibinfo
  {journal} {Phys. Rev. B}\ }\textbf {\bibinfo {volume} {96}},\ \bibinfo
  {pages} {035153} (\bibinfo {year} {2017})}\BibitemShut {NoStop}%
\bibitem [{\citenamefont {Schreiber}\ \emph {et~al.}(2015)\citenamefont
  {Schreiber}, \citenamefont {Hodgman}, \citenamefont {Bordia} \emph
  {et~al.}}]{30}%
  \BibitemOpen
  \bibfield  {author} {\bibinfo {author} {\bibfnamefont {Michael}\ \bibnamefont
  {Schreiber}}, \bibinfo {author} {\bibfnamefont {Sean~S.}\ \bibnamefont
  {Hodgman}}, \bibinfo {author} {\bibfnamefont {Pranjal}\ \bibnamefont
  {Bordia}},  \emph {et~al.},\ }\bibfield  {title} {\enquote {\bibinfo {title}
  {Observation of many-body localization of interacting fermions in a
  quasirandom optical lattice},}\ }\href
  {https://science.sciencemag.org/content/349/6250/842} {\bibfield  {journal}
  {\bibinfo  {journal} {Science}\ }\textbf {\bibinfo {volume} {349}},\ \bibinfo
  {pages} {842--845} (\bibinfo {year} {2015})}\BibitemShut {NoStop}%
\bibitem [{\citenamefont {Bardarson}\ \emph {et~al.}(2012)\citenamefont
  {Bardarson}, \citenamefont {Pollmann},\ and\ \citenamefont {Moore}}]{7}%
  \BibitemOpen
  \bibfield  {author} {\bibinfo {author} {\bibfnamefont {Jens~H.}\ \bibnamefont
  {Bardarson}}, \bibinfo {author} {\bibfnamefont {Frank}\ \bibnamefont
  {Pollmann}}, \ and\ \bibinfo {author} {\bibfnamefont {Joel~E.}\ \bibnamefont
  {Moore}},\ }\bibfield  {title} {\enquote {\bibinfo {title} {{Unbounded growth
  of entanglement in models of many-body localization}},}\ }\href {\doibase
  https://doi.org/10.1103/PhysRevLett.109.017202} {\bibfield  {journal}
  {\bibinfo  {journal} {Phys. Rev. Lett.}\ }\textbf {\bibinfo {volume} {109}},\
  \bibinfo {pages} {017202} (\bibinfo {year} {2012})}\BibitemShut {NoStop}%
\bibitem [{\citenamefont {Serbyn}\ \emph
  {et~al.}(2013{\natexlab{b}})\citenamefont {Serbyn}, \citenamefont {Papić},\
  and\ \citenamefont {Abanin}}]{8}%
  \BibitemOpen
  \bibfield  {author} {\bibinfo {author} {\bibfnamefont {Maksym}\ \bibnamefont
  {Serbyn}}, \bibinfo {author} {\bibfnamefont {Z.}~\bibnamefont {Papić}}, \
  and\ \bibinfo {author} {\bibfnamefont {Dmitry~A.}\ \bibnamefont {Abanin}},\
  }\bibfield  {title} {\enquote {\bibinfo {title} {Universal slow growth of
  entanglement in interacting strongly disordered systems},}\ }\href {\doibase
  https://doi.org/10.1103/PhysRevLett.110.260601} {\bibfield  {journal}
  {\bibinfo  {journal} {Phys. Rev. Lett.}\ }\textbf {\bibinfo {volume} {110}},\
  \bibinfo {pages} {260601} (\bibinfo {year} {2013}{\natexlab{b}})}\BibitemShut
  {NoStop}%
\bibitem [{\citenamefont {Lukin}\ \emph {et~al.}(2019)\citenamefont {Lukin},
  \citenamefont {Rispoli}, \citenamefont {Schittko} \emph {et~al.}}]{35}%
  \BibitemOpen
  \bibfield  {author} {\bibinfo {author} {\bibfnamefont {Alexander}\
  \bibnamefont {Lukin}}, \bibinfo {author} {\bibfnamefont {Matthew}\
  \bibnamefont {Rispoli}}, \bibinfo {author} {\bibfnamefont {Robert}\
  \bibnamefont {Schittko}},  \emph {et~al.},\ }\bibfield  {title} {\enquote
  {\bibinfo {title} {Probing entanglement in a many-body--localized system},}\
  }\href {https://science.sciencemag.org/content/364/6437/256} {\bibfield
  {journal} {\bibinfo  {journal} {Science}\ }\textbf {\bibinfo {volume}
  {364}},\ \bibinfo {pages} {256--260} (\bibinfo {year} {2019})}\BibitemShut
  {NoStop}%
\bibitem [{\citenamefont {Choi}\ \emph {et~al.}(2016)\citenamefont {Choi},
  \citenamefont {Hild}, \citenamefont {Zeiher} \emph {et~al.}}]{49}%
  \BibitemOpen
  \bibfield  {author} {\bibinfo {author} {\bibfnamefont {Jae-yoon}\
  \bibnamefont {Choi}}, \bibinfo {author} {\bibfnamefont {Sebastian}\
  \bibnamefont {Hild}}, \bibinfo {author} {\bibfnamefont {Johannes}\
  \bibnamefont {Zeiher}},  \emph {et~al.},\ }\bibfield  {title} {\enquote
  {\bibinfo {title} {Exploring the many-body localization transition in two
  dimensions},}\ }\href {https://www.science.org/doi/10.1126/science.aaf8834}
  {\bibfield  {journal} {\bibinfo  {journal} {Science}\ }\textbf {\bibinfo
  {volume} {352}},\ \bibinfo {pages} {1547} (\bibinfo {year}
  {2016})}\BibitemShut {NoStop}%
\bibitem [{\citenamefont {Guo}\ \emph {et~al.}(2021)\citenamefont {Guo},
  \citenamefont {Cheng}, \citenamefont {Sun} \emph {et~al.}}]{70}%
  \BibitemOpen
  \bibfield  {author} {\bibinfo {author} {\bibfnamefont {Qiujiang}\
  \bibnamefont {Guo}}, \bibinfo {author} {\bibfnamefont {Chen}\ \bibnamefont
  {Cheng}}, \bibinfo {author} {\bibfnamefont {Zheng-Hang}\ \bibnamefont {Sun}},
   \emph {et~al.},\ }\bibfield  {title} {\enquote {\bibinfo {title}
  {Observation of energy-resolved many-body localization},}\ }\href {\doibase
  https://doi.org/10.1038/s41567-020-1035-1} {\bibfield  {journal} {\bibinfo
  {journal} {Nat. Phys.}\ }\textbf {\bibinfo {volume} {17}},\ \bibinfo {pages}
  {234–239} (\bibinfo {year} {2021})}\BibitemShut {NoStop}%
\bibitem [{\citenamefont {Rispoli}\ \emph {et~al.}(2019)\citenamefont
  {Rispoli}, \citenamefont {Lukin}, \citenamefont {Schittko} \emph
  {et~al.}}]{73}%
  \BibitemOpen
  \bibfield  {author} {\bibinfo {author} {\bibfnamefont {Matthew}\ \bibnamefont
  {Rispoli}}, \bibinfo {author} {\bibfnamefont {Alexander}\ \bibnamefont
  {Lukin}}, \bibinfo {author} {\bibfnamefont {Robert}\ \bibnamefont
  {Schittko}},  \emph {et~al.},\ }\bibfield  {title} {\enquote {\bibinfo
  {title} {Quantum critical behaviour at the many-body localization
  transition},}\ }\href {\doibase https://doi.org/10.1038/s41586-019-1527-2}
  {\bibfield  {journal} {\bibinfo  {journal} {Nature}\ }\textbf {\bibinfo
  {volume} {573}},\ \bibinfo {pages} {385–389} (\bibinfo {year}
  {2019})}\BibitemShut {NoStop}%
\bibitem [{\citenamefont {L{\'e}onard}\ \emph {et~al.}(2023)\citenamefont
  {L{\'e}onard}, \citenamefont {Kim}, \citenamefont {Rispoli}, \citenamefont
  {Lukin}, \citenamefont {Schittko}, \citenamefont {Kwan}, \citenamefont
  {Demler}, \citenamefont {Sels},\ and\ \citenamefont
  {Greiner}}]{leonard2023probing}%
  \BibitemOpen
  \bibfield  {author} {\bibinfo {author} {\bibfnamefont {Julian}\ \bibnamefont
  {L{\'e}onard}}, \bibinfo {author} {\bibfnamefont {Sooshin}\ \bibnamefont
  {Kim}}, \bibinfo {author} {\bibfnamefont {Matthew}\ \bibnamefont {Rispoli}},
  \bibinfo {author} {\bibfnamefont {Alexander}\ \bibnamefont {Lukin}}, \bibinfo
  {author} {\bibfnamefont {Robert}\ \bibnamefont {Schittko}}, \bibinfo {author}
  {\bibfnamefont {Joyce}\ \bibnamefont {Kwan}}, \bibinfo {author}
  {\bibfnamefont {Eugene}\ \bibnamefont {Demler}}, \bibinfo {author}
  {\bibfnamefont {Dries}\ \bibnamefont {Sels}}, \ and\ \bibinfo {author}
  {\bibfnamefont {Markus}\ \bibnamefont {Greiner}},\ }\bibfield  {title}
  {\enquote {\bibinfo {title} {Probing the onset of quantum avalanches in a
  many-body localized system},}\ }\href
  {https://www.nature.com/articles/s41567-022-01887-3} {\bibfield  {journal}
  {\bibinfo  {journal} {Nat. Phys.}\ }\textbf {\bibinfo {volume} {19}},\
  \bibinfo {pages} {481--485} (\bibinfo {year} {2023})}\BibitemShut {NoStop}%
\bibitem [{\citenamefont {Zhang}\ and\ \citenamefont {Dong}(2010)}]{28}%
  \BibitemOpen
  \bibfield  {author} {\bibinfo {author} {\bibfnamefont {J.~M.}\ \bibnamefont
  {Zhang}}\ and\ \bibinfo {author} {\bibfnamefont {R.~X.}\ \bibnamefont
  {Dong}},\ }\bibfield  {title} {\enquote {\bibinfo {title} {Exact
  diagonalization: the bose--hubbard model as an example},}\ }\href
  {https://iopscience.iop.org/article/10.1088/0143-0807/31/3/016/meta}
  {\bibfield  {journal} {\bibinfo  {journal} {Eur. J. Phys.}\ }\textbf
  {\bibinfo {volume} {31}},\ \bibinfo {pages} {3} (\bibinfo {year}
  {2010})}\BibitemShut {NoStop}%
\bibitem [{\citenamefont {Paeckela}\ \emph {et~al.}(2019)\citenamefont
  {Paeckela}, \citenamefont {Köhlera}, \citenamefont {Swobodab} \emph
  {et~al.}}]{37}%
  \BibitemOpen
  \bibfield  {author} {\bibinfo {author} {\bibfnamefont {Sebastian}\
  \bibnamefont {Paeckela}}, \bibinfo {author} {\bibfnamefont {Thomas}\
  \bibnamefont {Köhlera}}, \bibinfo {author} {\bibfnamefont {Andreas}\
  \bibnamefont {Swobodab}},  \emph {et~al.},\ }\bibfield  {title} {\enquote
  {\bibinfo {title} {Time-evolution methods for matrix-product states},}\
  }\href {\doibase https://doi.org/10.1016/j.aop.2019.167998} {\bibfield
  {journal} {\bibinfo  {journal} {Annals of Physics}\ }\textbf {\bibinfo
  {volume} {411}},\ \bibinfo {pages} {167998} (\bibinfo {year}
  {2019})}\BibitemShut {NoStop}%
\bibitem [{\citenamefont {Vidal}(2004)}]{38}%
  \BibitemOpen
  \bibfield  {author} {\bibinfo {author} {\bibfnamefont {Guifre}\ \bibnamefont
  {Vidal}},\ }\bibfield  {title} {\enquote {\bibinfo {title} {Efficient
  simulation of one-dimensional quantum many-body systems},}\ }\href {\doibase
  https://doi.org/10.1103/PhysRevLett.93.040502} {\bibfield  {journal}
  {\bibinfo  {journal} {Phys. Rev. Lett.}\ }\textbf {\bibinfo {volume} {93}},\
  \bibinfo {pages} {040502} (\bibinfo {year} {2004})}\BibitemShut {NoStop}%
\bibitem [{\citenamefont {Kennett}(2013)}]{39}%
  \BibitemOpen
  \bibfield  {author} {\bibinfo {author} {\bibfnamefont {Malcolm~P.}\
  \bibnamefont {Kennett}},\ }\bibfield  {title} {\enquote {\bibinfo {title}
  {Out-of-equilibrium dynamics of the bose-hubbard model},}\ }\href {\doibase
  http://dx.doi.org/10.1155/2013/393616} {\bibfield  {journal} {\bibinfo
  {journal} {ISRN Condensed Matter Physics}\ }\textbf {\bibinfo {volume}
  {393616}},\ \bibinfo {pages} {39} (\bibinfo {year} {2013})}\BibitemShut
  {NoStop}%
\bibitem [{\citenamefont {Sengupta}\ and\ \citenamefont {Haas}(2007)}]{40}%
  \BibitemOpen
  \bibfield  {author} {\bibinfo {author} {\bibfnamefont {Pinaki}\ \bibnamefont
  {Sengupta}}\ and\ \bibinfo {author} {\bibfnamefont {Stephan}\ \bibnamefont
  {Haas}},\ }\bibfield  {title} {\enquote {\bibinfo {title} {Quantum glass
  phases in the disordered bose-hubbard model},}\ }\href {\doibase
  https://doi.org/10.1103/PhysRevLett.99.050403} {\bibfield  {journal}
  {\bibinfo  {journal} {Phys. Rev. Lett.}\ }\textbf {\bibinfo {volume} {99}},\
  \bibinfo {pages} {050403} (\bibinfo {year} {2007})}\BibitemShut {NoStop}%
\bibitem [{\citenamefont {Gurarie}\ \emph {et~al.}(2009)\citenamefont
  {Gurarie}, \citenamefont {Pollet}, \citenamefont {Prokof’ev} \emph
  {et~al.}}]{41}%
  \BibitemOpen
  \bibfield  {author} {\bibinfo {author} {\bibfnamefont {V.}~\bibnamefont
  {Gurarie}}, \bibinfo {author} {\bibfnamefont {L.}~\bibnamefont {Pollet}},
  \bibinfo {author} {\bibfnamefont {N.~V.}\ \bibnamefont {Prokof’ev}},  \emph
  {et~al.},\ }\bibfield  {title} {\enquote {\bibinfo {title} {Phase diagram of
  the disordered bose-hubbard model},}\ }\href {\doibase
  https://doi.org/10.1103/PhysRevB.80.214519} {\bibfield  {journal} {\bibinfo
  {journal} {Phys. Rev. B}\ }\textbf {\bibinfo {volume} {80}},\ \bibinfo
  {pages} {214519} (\bibinfo {year} {2009})}\BibitemShut {NoStop}%
\bibitem [{\citenamefont {Pollet}\ \emph {et~al.}(2009)\citenamefont {Pollet},
  \citenamefont {Prokof’ev}, \citenamefont {Svistunov} \emph {et~al.}}]{42}%
  \BibitemOpen
  \bibfield  {author} {\bibinfo {author} {\bibfnamefont {L.}~\bibnamefont
  {Pollet}}, \bibinfo {author} {\bibfnamefont {N.~V.}\ \bibnamefont
  {Prokof’ev}}, \bibinfo {author} {\bibfnamefont {B.~V.}\ \bibnamefont
  {Svistunov}},  \emph {et~al.},\ }\bibfield  {title} {\enquote {\bibinfo
  {title} {Absence of a direct superfluid to mott insulator transition in
  disordered bose systems},}\ }\href {\doibase
  https://doi.org/10.1103/PhysRevLett.103.140402} {\bibfield  {journal}
  {\bibinfo  {journal} {Phys. Rev. Lett.}\ }\textbf {\bibinfo {volume} {103}},\
  \bibinfo {pages} {140402} (\bibinfo {year} {2009})}\BibitemShut {NoStop}%
\bibitem [{\citenamefont {Gerster}\ \emph {et~al.}(2016)\citenamefont
  {Gerster}, \citenamefont {Rizzi}, \citenamefont {Tschirsich} \emph
  {et~al.}}]{43}%
  \BibitemOpen
  \bibfield  {author} {\bibinfo {author} {\bibfnamefont {M.}~\bibnamefont
  {Gerster}}, \bibinfo {author} {\bibfnamefont {M.}~\bibnamefont {Rizzi}},
  \bibinfo {author} {\bibfnamefont {F.}~\bibnamefont {Tschirsich}},  \emph
  {et~al.},\ }\bibfield  {title} {\enquote {\bibinfo {title} {Superfluid
  density and quasi-long-range order in the one-dimensional disordered
  bose--hubbard model},}\ }\href
  {https://iopscience.iop.org/article/10.1088/1367-2630/18/1/015015} {\bibfield
   {journal} {\bibinfo  {journal} {New J. Phys.}\ }\textbf {\bibinfo {volume}
  {18}},\ \bibinfo {pages} {015015} (\bibinfo {year} {2016})}\BibitemShut
  {NoStop}%
\bibitem [{\citenamefont {Hu}\ \emph {et~al.}(2009)\citenamefont {Hu},
  \citenamefont {Wen}, \citenamefont {Yu}, \citenamefont {Normand},\ and\
  \citenamefont {Wang}}]{50}%
  \BibitemOpen
  \bibfield  {author} {\bibinfo {author} {\bibfnamefont {Shijie}\ \bibnamefont
  {Hu}}, \bibinfo {author} {\bibfnamefont {Yuchuan}\ \bibnamefont {Wen}},
  \bibinfo {author} {\bibfnamefont {Yue}\ \bibnamefont {Yu}}, \bibinfo {author}
  {\bibfnamefont {B.}~\bibnamefont {Normand}}, \ and\ \bibinfo {author}
  {\bibfnamefont {Xiaoqun}\ \bibnamefont {Wang}},\ }\bibfield  {title}
  {\enquote {\bibinfo {title} {Quantized squeezing and even-odd asymmetry of
  trapped bosons},}\ }\href {\doibase
  https://doi.org/10.1103/PhysRevA.80.063624} {\bibfield  {journal} {\bibinfo
  {journal} {Phys. Rev. A}\ }\textbf {\bibinfo {volume} {80}},\ \bibinfo
  {pages} {063624} (\bibinfo {year} {2009})}\BibitemShut {NoStop}%
\bibitem [{\citenamefont {{\v{S}}untajs}\ \emph
  {et~al.}(2020{\natexlab{a}})\citenamefont {{\v{S}}untajs}, \citenamefont
  {Bon{\v{c}}a}, \citenamefont {Prosen},\ and\ \citenamefont
  {Vidmar}}]{vsuntajs2020quantum}%
  \BibitemOpen
  \bibfield  {author} {\bibinfo {author} {\bibfnamefont {Jan}\ \bibnamefont
  {{\v{S}}untajs}}, \bibinfo {author} {\bibfnamefont {Janez}\ \bibnamefont
  {Bon{\v{c}}a}}, \bibinfo {author} {\bibfnamefont {Toma{\v{z}}}\ \bibnamefont
  {Prosen}}, \ and\ \bibinfo {author} {\bibfnamefont {Lev}\ \bibnamefont
  {Vidmar}},\ }\bibfield  {title} {\enquote {\bibinfo {title} {Quantum chaos
  challenges many-body localization},}\ }\href {\doibase
  https://doi.org/10.1103/PhysRevE.102.062144} {\bibfield  {journal} {\bibinfo
  {journal} {Phys. Rev. E}\ }\textbf {\bibinfo {volume} {102}},\ \bibinfo
  {pages} {062144} (\bibinfo {year} {2020}{\natexlab{a}})}\BibitemShut
  {NoStop}%
\bibitem [{\citenamefont {Morningstar}\ \emph {et~al.}(2022)\citenamefont
  {Morningstar}, \citenamefont {Colmenarez}, \citenamefont {Khemani},
  \citenamefont {Luitz},\ and\ \citenamefont
  {Huse}}]{morningstar2022avalanches}%
  \BibitemOpen
  \bibfield  {author} {\bibinfo {author} {\bibfnamefont {Alan}\ \bibnamefont
  {Morningstar}}, \bibinfo {author} {\bibfnamefont {Luis}\ \bibnamefont
  {Colmenarez}}, \bibinfo {author} {\bibfnamefont {Vedika}\ \bibnamefont
  {Khemani}}, \bibinfo {author} {\bibfnamefont {David~J}\ \bibnamefont
  {Luitz}}, \ and\ \bibinfo {author} {\bibfnamefont {David~A}\ \bibnamefont
  {Huse}},\ }\bibfield  {title} {\enquote {\bibinfo {title} {Avalanches and
  many-body resonances in many-body localized systems},}\ }\href {\doibase
  https://doi.org/10.1103/PhysRevB.105.174205} {\bibfield  {journal} {\bibinfo
  {journal} {Phys. Rev. B}\ }\textbf {\bibinfo {volume} {105}},\ \bibinfo
  {pages} {174205} (\bibinfo {year} {2022})}\BibitemShut {NoStop}%
\bibitem [{\citenamefont {Widom}(2017)}]{68}%
  \BibitemOpen
  \bibfield  {author} {\bibinfo {author} {\bibfnamefont {B.}~\bibnamefont
  {Widom}},\ }\bibfield  {title} {\enquote {\bibinfo {title} {Equation of state
  in the neighborhood of the critical point},}\ }\href {\doibase
  https://doi.org/10.1063/1.1696618} {\bibfield  {journal} {\bibinfo  {journal}
  {J. Chem. Phys.}\ }\textbf {\bibinfo {volume} {43}},\ \bibinfo {pages} {3892}
  (\bibinfo {year} {2017})}\BibitemShut {NoStop}%
\bibitem [{\citenamefont {Vojta}(2013)}]{66}%
  \BibitemOpen
  \bibfield  {author} {\bibinfo {author} {\bibfnamefont {Thomas}\ \bibnamefont
  {Vojta}},\ }\bibfield  {title} {\enquote {\bibinfo {title} {Phases and phase
  transitions in disordered quantum systems},}\ }\href {\doibase
  https://doi.org/10.1063/1.4818403} {\bibfield  {journal} {\bibinfo  {journal}
  {AIP Conf. Proc.}\ }\textbf {\bibinfo {volume} {1550}},\ \bibinfo {pages}
  {188--247} (\bibinfo {year} {2013})}\BibitemShut {NoStop}%
\bibitem [{\citenamefont {Orell}\ \emph
  {et~al.}(2019{\natexlab{a}})\citenamefont {Orell}, \citenamefont
  {Michailidis}, \citenamefont {Serbyn},\ and\ \citenamefont {Silveri}}]{86}%
  \BibitemOpen
  \bibfield  {author} {\bibinfo {author} {\bibfnamefont {Tuure}\ \bibnamefont
  {Orell}}, \bibinfo {author} {\bibfnamefont {Alexios~A.}\ \bibnamefont
  {Michailidis}}, \bibinfo {author} {\bibfnamefont {Maksym}\ \bibnamefont
  {Serbyn}}, \ and\ \bibinfo {author} {\bibfnamefont {Matti}\ \bibnamefont
  {Silveri}},\ }\bibfield  {title} {\enquote {\bibinfo {title} {Probing the
  many-body localization phase transition with superconducting circuits},}\
  }\href {\doibase https://doi.org/10.1103/PhysRevB.100.134504} {\bibfield
  {journal} {\bibinfo  {journal} {Phys. Rev. B}\ }\textbf {\bibinfo {volume}
  {100}},\ \bibinfo {pages} {134504} (\bibinfo {year}
  {2019}{\natexlab{a}})}\BibitemShut {NoStop}%
\bibitem [{\citenamefont {Zhang}\ and\ \citenamefont {Yao}(2018)}]{65}%
  \BibitemOpen
  \bibfield  {author} {\bibinfo {author} {\bibfnamefont {Shi-Xin}\ \bibnamefont
  {Zhang}}\ and\ \bibinfo {author} {\bibfnamefont {Hong}\ \bibnamefont {Yao}},\
  }\bibfield  {title} {\enquote {\bibinfo {title} {Universal properties of
  many-body localization transitions in quasiperiodic systems},}\ }\href
  {\doibase https://doi.org/10.1103/PhysRevLett.121.206601} {\bibfield
  {journal} {\bibinfo  {journal} {Phys. Rev. Lett.}\ }\textbf {\bibinfo
  {volume} {121}},\ \bibinfo {pages} {206601} (\bibinfo {year}
  {2018})}\BibitemShut {NoStop}%
\bibitem [{\citenamefont {Khemani}\ \emph
  {et~al.}(2017{\natexlab{a}})\citenamefont {Khemani}, \citenamefont {Sheng},\
  and\ \citenamefont {Huse}}]{88}%
  \BibitemOpen
  \bibfield  {author} {\bibinfo {author} {\bibfnamefont {Vedika}\ \bibnamefont
  {Khemani}}, \bibinfo {author} {\bibfnamefont {D.~N.}\ \bibnamefont {Sheng}},
  \ and\ \bibinfo {author} {\bibfnamefont {David~A.}\ \bibnamefont {Huse}},\
  }\bibfield  {title} {\enquote {\bibinfo {title} {Two universality classes for
  the many-body localization transition},}\ }\href
  {https://doi.org/10.1103/PhysRevLett.119.075702} {\bibfield  {journal}
  {\bibinfo  {journal} {Phys. Rev. Lett.}\ }\textbf {\bibinfo {volume} {119}},\
  \bibinfo {pages} {075702} (\bibinfo {year} {2017}{\natexlab{a}})}\BibitemShut
  {NoStop}%
\bibitem [{\citenamefont {{\v{S}}untajs}\ \emph
  {et~al.}(2020{\natexlab{b}})\citenamefont {{\v{S}}untajs}, \citenamefont
  {Bon{\v{c}}a}, \citenamefont {Prosen},\ and\ \citenamefont
  {Vidmar}}]{vsuntajs2020ergodicity}%
  \BibitemOpen
  \bibfield  {author} {\bibinfo {author} {\bibfnamefont {Jan}\ \bibnamefont
  {{\v{S}}untajs}}, \bibinfo {author} {\bibfnamefont {Janez}\ \bibnamefont
  {Bon{\v{c}}a}}, \bibinfo {author} {\bibfnamefont {Toma{\v{z}}}\ \bibnamefont
  {Prosen}}, \ and\ \bibinfo {author} {\bibfnamefont {Lev}\ \bibnamefont
  {Vidmar}},\ }\bibfield  {title} {\enquote {\bibinfo {title} {Ergodicity
  breaking transition in finite disordered spin chains},}\ }\href {\doibase
  https://doi.org/10.1103/PhysRevB.102.064207} {\bibfield  {journal} {\bibinfo
  {journal} {Phys. Rev. B}\ }\textbf {\bibinfo {volume} {102}},\ \bibinfo
  {pages} {064207} (\bibinfo {year} {2020}{\natexlab{b}})}\BibitemShut
  {NoStop}%
\bibitem [{\citenamefont {Sierant}\ \emph
  {et~al.}(2017{\natexlab{a}})\citenamefont {Sierant}, \citenamefont
  {Delande},\ and\ \citenamefont {Zakrzewski}}]{76}%
  \BibitemOpen
  \bibfield  {author} {\bibinfo {author} {\bibfnamefont {Piotr}\ \bibnamefont
  {Sierant}}, \bibinfo {author} {\bibfnamefont {Dominique}\ \bibnamefont
  {Delande}}, \ and\ \bibinfo {author} {\bibfnamefont {Jakub}\ \bibnamefont
  {Zakrzewski}},\ }\bibfield  {title} {\enquote {\bibinfo {title} {Many-body
  localization for randomly interacting bosons},}\ }\href {\doibase
  10.12693/APhysPolA.132.1707} {\bibfield  {journal} {\bibinfo  {journal} {Acta
  Physica Polonica A}\ }\textbf {\bibinfo {volume} {132}},\ \bibinfo {pages}
  {1707--1712} (\bibinfo {year} {2017}{\natexlab{a}})}\BibitemShut {NoStop}%
\bibitem [{\citenamefont {Sierant}\ \emph
  {et~al.}(2017{\natexlab{b}})\citenamefont {Sierant}, \citenamefont
  {Delande},\ and\ \citenamefont {Zakrzewski}}]{77}%
  \BibitemOpen
  \bibfield  {author} {\bibinfo {author} {\bibfnamefont {Piotr}\ \bibnamefont
  {Sierant}}, \bibinfo {author} {\bibfnamefont {Dominique}\ \bibnamefont
  {Delande}}, \ and\ \bibinfo {author} {\bibfnamefont {Jakub}\ \bibnamefont
  {Zakrzewski}},\ }\bibfield  {title} {\enquote {\bibinfo {title} {Many-body
  localization due to random interactions},}\ }\href {\doibase
  https://doi.org/10.1103/PhysRevA.95.021601} {\bibfield  {journal} {\bibinfo
  {journal} {Phys. Rev. A}\ }\textbf {\bibinfo {volume} {95}},\ \bibinfo
  {pages} {021601(R)} (\bibinfo {year} {2017}{\natexlab{b}})}\BibitemShut
  {NoStop}%
\bibitem [{\citenamefont {Orell}\ \emph
  {et~al.}(2019{\natexlab{b}})\citenamefont {Orell}, \citenamefont
  {Michailidis}, \citenamefont {Serbyn},\ and\ \citenamefont
  {Silveri}}]{orell2019probing}%
  \BibitemOpen
  \bibfield  {author} {\bibinfo {author} {\bibfnamefont {Tuure}\ \bibnamefont
  {Orell}}, \bibinfo {author} {\bibfnamefont {Alexios~A}\ \bibnamefont
  {Michailidis}}, \bibinfo {author} {\bibfnamefont {Maksym}\ \bibnamefont
  {Serbyn}}, \ and\ \bibinfo {author} {\bibfnamefont {Matti}\ \bibnamefont
  {Silveri}},\ }\bibfield  {title} {\enquote {\bibinfo {title} {Probing the
  many-body localization phase transition with superconducting circuits},}\
  }\href {\doibase https://doi.org/10.1103/PhysRevB.100.134504} {\bibfield
  {journal} {\bibinfo  {journal} {Phys. Rev. B}\ }\textbf {\bibinfo {volume}
  {100}},\ \bibinfo {pages} {134504} (\bibinfo {year}
  {2019}{\natexlab{b}})}\BibitemShut {NoStop}%
\bibitem [{\citenamefont {Atas}\ \emph {et~al.}(2013)\citenamefont {Atas},
  \citenamefont {Bogomolny}, \citenamefont {Giraud},\ and\ \citenamefont
  {Roux}}]{81}%
  \BibitemOpen
  \bibfield  {author} {\bibinfo {author} {\bibfnamefont {Y.~Y.}\ \bibnamefont
  {Atas}}, \bibinfo {author} {\bibfnamefont {E.}~\bibnamefont {Bogomolny}},
  \bibinfo {author} {\bibfnamefont {O.}~\bibnamefont {Giraud}}, \ and\ \bibinfo
  {author} {\bibfnamefont {G.}~\bibnamefont {Roux}},\ }\bibfield  {title}
  {\enquote {\bibinfo {title} {Distribution of the ratio of consecutive level
  spacings in random matrix ensembles},}\ }\href {\doibase
  https://doi.org/10.1103/PhysRevLett.110.084101} {\bibfield  {journal}
  {\bibinfo  {journal} {Phys. Rev. Lett.}\ }\textbf {\bibinfo {volume} {110}},\
  \bibinfo {pages} {084101} (\bibinfo {year} {2013})}\BibitemShut {NoStop}%
\bibitem [{\citenamefont {Chen}\ \emph {et~al.}()\citenamefont {Chen},
  \citenamefont {Chen},\ and\ \citenamefont {Wang}}]{84}%
  \BibitemOpen
  \bibfield  {author} {\bibinfo {author} {\bibfnamefont {Jie}\ \bibnamefont
  {Chen}}, \bibinfo {author} {\bibfnamefont {Chun}\ \bibnamefont {Chen}}, \
  and\ \bibinfo {author} {\bibfnamefont {Xiaoqun}\ \bibnamefont {Wang}},\
  }\href {https://arxiv.org/abs/2303.14825} {\enquote {\bibinfo {title}
  {{Energy- and Symmetry-Resolved Entanglement Dynamics in Disordered
  Bose-Hubbard Chain}},}\ }\Eprint {http://arxiv.org/abs/2303.14825}
  {arXiv:2303.14825} \BibitemShut {NoStop}%
\bibitem [{\citenamefont {Luitz}\ and\ \citenamefont {Lev}(2020)}]{79}%
  \BibitemOpen
  \bibfield  {author} {\bibinfo {author} {\bibfnamefont {David~J.}\
  \bibnamefont {Luitz}}\ and\ \bibinfo {author} {\bibfnamefont {Yevgeny~Bar}\
  \bibnamefont {Lev}},\ }\bibfield  {title} {\enquote {\bibinfo {title}
  {Absence of slow particle transport in the many-body localized phase},}\
  }\href {\doibase https://doi.org/10.1103/PhysRevB.102.100202} {\bibfield
  {journal} {\bibinfo  {journal} {Phys. Rev. B}\ }\textbf {\bibinfo {volume}
  {102}},\ \bibinfo {pages} {100202(R)} (\bibinfo {year} {2020})}\BibitemShut
  {NoStop}%
\bibitem [{\citenamefont {Kiefer-Emmanouilidis}\ \emph
  {et~al.}(2020)\citenamefont {Kiefer-Emmanouilidis}, \citenamefont {Unanyan},
  \citenamefont {Fleischhauer},\ and\ \citenamefont {Sirker}}]{80}%
  \BibitemOpen
  \bibfield  {author} {\bibinfo {author} {\bibfnamefont {Maximilian}\
  \bibnamefont {Kiefer-Emmanouilidis}}, \bibinfo {author} {\bibfnamefont
  {Razmik}\ \bibnamefont {Unanyan}}, \bibinfo {author} {\bibfnamefont
  {Michael}\ \bibnamefont {Fleischhauer}}, \ and\ \bibinfo {author}
  {\bibfnamefont {Jesko}\ \bibnamefont {Sirker}},\ }\bibfield  {title}
  {\enquote {\bibinfo {title} {Evidence for unbounded growth of the number
  entropy in many-body localized phases},}\ }\href {\doibase
  https://doi.org/10.1103/PhysRevLett.124.243601} {\bibfield  {journal}
  {\bibinfo  {journal} {Phys. Rev. Lett.}\ }\textbf {\bibinfo {volume} {124}},\
  \bibinfo {pages} {243601} (\bibinfo {year} {2020})}\BibitemShut {NoStop}%
\bibitem [{\citenamefont {Luitz}\ \emph {et~al.}(2016)\citenamefont {Luitz},
  \citenamefont {Laflorencie},\ and\ \citenamefont {Alet}}]{26}%
  \BibitemOpen
  \bibfield  {author} {\bibinfo {author} {\bibfnamefont {David~J.}\
  \bibnamefont {Luitz}}, \bibinfo {author} {\bibfnamefont {Nicolas}\
  \bibnamefont {Laflorencie}}, \ and\ \bibinfo {author} {\bibfnamefont
  {Fabien}\ \bibnamefont {Alet}},\ }\bibfield  {title} {\enquote {\bibinfo
  {title} {Extended slow dynamical regime close to the many-body localization
  transition},}\ }\href {\doibase https://doi.org/10.1103/PhysRevB.93.060201}
  {\bibfield  {journal} {\bibinfo  {journal} {Phys. Rev. B}\ }\textbf {\bibinfo
  {volume} {93}},\ \bibinfo {pages} {060201(R)} (\bibinfo {year}
  {2016})}\BibitemShut {NoStop}%
\bibitem [{\citenamefont {Kohlert}\ \emph {et~al.}(2019)\citenamefont
  {Kohlert}, \citenamefont {Scherg}, \citenamefont {Li} \emph {et~al.}}]{69}%
  \BibitemOpen
  \bibfield  {author} {\bibinfo {author} {\bibfnamefont {Thomas}\ \bibnamefont
  {Kohlert}}, \bibinfo {author} {\bibfnamefont {Sebastian}\ \bibnamefont
  {Scherg}}, \bibinfo {author} {\bibfnamefont {Xiao}\ \bibnamefont {Li}},
  \emph {et~al.},\ }\bibfield  {title} {\enquote {\bibinfo {title} {Observation
  of many-body localization in a one-dimensional system with a single-particle
  mobility edge},}\ }\href {\doibase
  https://doi.org/10.1103/PhysRevLett.122.170403} {\bibfield  {journal}
  {\bibinfo  {journal} {Phys. Rev. Lett.}\ }\textbf {\bibinfo {volume} {122}},\
  \bibinfo {pages} {170403} (\bibinfo {year} {2019})}\BibitemShut {NoStop}%
\bibitem [{\citenamefont {Agarwal}\ \emph {et~al.}(2017)\citenamefont
  {Agarwal}, \citenamefont {Altman}, \citenamefont {Demler} \emph
  {et~al.}}]{67}%
  \BibitemOpen
  \bibfield  {author} {\bibinfo {author} {\bibfnamefont {Kartiek}\ \bibnamefont
  {Agarwal}}, \bibinfo {author} {\bibfnamefont {Ehud}\ \bibnamefont {Altman}},
  \bibinfo {author} {\bibfnamefont {Eugene}\ \bibnamefont {Demler}},  \emph
  {et~al.},\ }\bibfield  {title} {\enquote {\bibinfo {title} {Rare-region
  effects and dynamics near the many-body localization transition},}\ }\href
  {\doibase https://doi.org/10.1002/andp.201600326} {\bibfield  {journal}
  {\bibinfo  {journal} {Ann. Phys. (Berlin)}\ }\textbf {\bibinfo {volume}
  {529}},\ \bibinfo {pages} {7--1600326} (\bibinfo {year} {2017})}\BibitemShut
  {NoStop}%
\bibitem [{\citenamefont {Khemani}\ \emph
  {et~al.}(2017{\natexlab{b}})\citenamefont {Khemani}, \citenamefont {Sheng},\
  and\ \citenamefont {Huse}}]{63}%
  \BibitemOpen
  \bibfield  {author} {\bibinfo {author} {\bibfnamefont {Vedika}\ \bibnamefont
  {Khemani}}, \bibinfo {author} {\bibfnamefont {D.~N.}\ \bibnamefont {Sheng}},
  \ and\ \bibinfo {author} {\bibfnamefont {David~A.}\ \bibnamefont {Huse}},\
  }\bibfield  {title} {\enquote {\bibinfo {title} {Two universality classes for
  the many-body localization transition},}\ }\href {\doibase
  https://doi.org/10.1103/PhysRevLett.119.075702} {\bibfield  {journal}
  {\bibinfo  {journal} {Phys. Rev. Lett.}\ }\textbf {\bibinfo {volume} {119}},\
  \bibinfo {pages} {075702} (\bibinfo {year} {2017}{\natexlab{b}})}\BibitemShut
  {NoStop}%
\bibitem [{\citenamefont {Khemani}\ \emph
  {et~al.}(2017{\natexlab{c}})\citenamefont {Khemani}, \citenamefont {Lim},
  \citenamefont {Sheng},\ and\ \citenamefont {Huse}}]{64}%
  \BibitemOpen
  \bibfield  {author} {\bibinfo {author} {\bibfnamefont {Vedika}\ \bibnamefont
  {Khemani}}, \bibinfo {author} {\bibfnamefont {S.~P.}\ \bibnamefont {Lim}},
  \bibinfo {author} {\bibfnamefont {D.~N.}\ \bibnamefont {Sheng}}, \ and\
  \bibinfo {author} {\bibfnamefont {David~A.}\ \bibnamefont {Huse}},\
  }\bibfield  {title} {\enquote {\bibinfo {title} {Critical properties of the
  many-body localization transition},}\ }\href {\doibase
  https://doi.org/10.1103/PhysRevX.7.021013} {\bibfield  {journal} {\bibinfo
  {journal} {Phys. Rev. X}\ }\textbf {\bibinfo {volume} {7}},\ \bibinfo {pages}
  {021013} (\bibinfo {year} {2017}{\natexlab{c}})}\BibitemShut {NoStop}%
\bibitem [{\citenamefont {Schierenberg}\ \emph {et~al.}(2012)\citenamefont
  {Schierenberg}, \citenamefont {Bruckmann},\ and\ \citenamefont
  {Wettig}}]{schierenberg2012wigner}%
  \BibitemOpen
  \bibfield  {author} {\bibinfo {author} {\bibfnamefont {Sebastian}\
  \bibnamefont {Schierenberg}}, \bibinfo {author} {\bibfnamefont {Falk}\
  \bibnamefont {Bruckmann}}, \ and\ \bibinfo {author} {\bibfnamefont {Tilo}\
  \bibnamefont {Wettig}},\ }\bibfield  {title} {\enquote {\bibinfo {title}
  {Wigner surmise for mixed symmetry classes in random matrix theory},}\ }\href
  {\doibase https://doi.org/10.1103/PhysRevE.85.061130} {\bibfield  {journal}
  {\bibinfo  {journal} {Phys. Rev. E}\ }\textbf {\bibinfo {volume} {85}},\
  \bibinfo {pages} {061130} (\bibinfo {year} {2012})}\BibitemShut {NoStop}%
\bibitem [{\citenamefont {Geraedts}\ \emph {et~al.}(2016)\citenamefont
  {Geraedts}, \citenamefont {Nandkishore},\ and\ \citenamefont
  {Regnault}}]{33}%
  \BibitemOpen
  \bibfield  {author} {\bibinfo {author} {\bibfnamefont {Scott~D.}\
  \bibnamefont {Geraedts}}, \bibinfo {author} {\bibfnamefont {Rahul}\
  \bibnamefont {Nandkishore}}, \ and\ \bibinfo {author} {\bibfnamefont
  {Nicolas}\ \bibnamefont {Regnault}},\ }\bibfield  {title} {\enquote {\bibinfo
  {title} {Many-body localization and thermalization: Insights from the
  entanglement spectrum},}\ }\href {\doibase
  https://doi.org/10.1103/PhysRevB.93.174202} {\bibfield  {journal} {\bibinfo
  {journal} {Phys. Rev. B}\ }\textbf {\bibinfo {volume} {93}},\ \bibinfo
  {pages} {174202} (\bibinfo {year} {2016})}\BibitemShut {NoStop}%
\end{thebibliography}%

\end{document}